\documentclass[cits]{PoS}
\usepackage{amsmath}
\usepackage[utf8]{inputenc}

\def\gev{\,\text{Ge\hspace{-0.1em}V}}
\def\mev{\,\text{Me\hspace{-0.1em}V}}
\providecommand{\abs}[1]{\lvert#1\rvert}    
\makeatletter
     {\bgroup\raggedright\small\section*{\refname
        \@mkboth{\MakeUppercase\refname}{\MakeUppercase\refname}}%
      \list{\name{bib\@arabic\c@enumiv}
            \@biblabel{\@arabic\c@enumiv}}%
           {\settowidth\labelwidth{\@biblabel{#1}}%
            \setlength\itemsep{0pt}
            \setlength\parskip{0pt}
            \setlength\parsep{0pt}
            \setlength\partopsep{0pt}
            \leftmargin\labelwidth
            \advance\leftmargin\labelsep
            \@openbib@code
            \usecounter{enumiv}%
            \let\p@enumiv\@empty
            \renewcommand\theenumiv{\@arabic\c@enumiv}}%
      \sloppy\clubpenalty4000\widowpenalty4000%
      \sfcode`\.\@m}
     {\def\@noitemerr
       {\@latex@warning{Empty `thebibliography' environment}}%
      \endlist\egroup}
\makeatother

\definecolor{darkcyan}{rgb}{0,0.5451,0.5451}
\definecolor{cyan}{rgb}{0.596,0.9611,0.961}
\definecolor{darkblue}{rgb}{0.0,0.0,0.5451}
\definecolor{lightblue}{rgb}{0.541,0.541,1}
\definecolor{pink}{rgb}{1.0,0.2,0.8}
\definecolor{lightpink}{rgb}{1.0,0.7,0.925}
\definecolor{darkred}{rgb}{0.7804,0.0,0.0}
\definecolor{lightred}{rgb}{1.0,0.478,0.478}
\definecolor{darkgreen}{rgb}{0, 0.388, 0}
\definecolor{lightgreen}{rgb}{0, 0.78, 0.388}
\definecolor{purple}{rgb}{0.4980,0, 0.4980}
\definecolor{brightergreen}{rgb}{0.35, 1.0, 0}
\definecolor{brightgreen}{rgb}{0.2, 0.4, 0}
\definecolor{lightpurple}{rgb}{1.0,0.3, 0.50}
\definecolor{purple}{rgb}{0.4980,0.0, 0.4980}
\definecolor{lightviolet}{rgb}{0.8,0.3, 0.8}
\definecolor{violet}{rgb}{0.3980,0., 0.3980}
\definecolor{gray}{rgb}{0.60,0.60, 0.60}
\definecolor{orange}{rgb}{1.0, 0.5, 0}
\definecolor{YellowOrange}{rgb}{1.0, 0.72, 0}
\definecolor{lightgray}{rgb}{0.6,0.6,0.6}
\title{Lattice QCD \newline (focus on Charm and Beauty form factors, $R(D^*)$, \newline $b$- \& $c$-quark masses)}

\ShortTitle{Lattice QCD}

\author{\speaker{Oliver Witzel}\\
        Department of Physics, University of Colorado Boulder, Boulder, CO 80303, USA\\
        E-mail: \email{Oliver.Witzel@colorado.edu}}

\abstract{We present an overview of state of the art lattice quantum chromodynamcis calculations for heavy-light quantities. Special focus is given to the calculation of form factors for semi-leptonic decays of $B_{(s)}$ and $D$ mesons, the extraction of the Cabibbo-Kobayashi-Maskawa matrix elements $|V_{ub}|$ and $|V_{cb}|$ as well as the determination of $R(D^{(*)})$ testing the universality of lepton flavors in $b\to c$ transitions. In addition we report on the determination of $b$ and $c$ quark masses as well as on neutral $B_{(s)}$ meson mixing. Recent results are summarized and new developments highlighted.}

\FullConference{18th International Conference on B-Physics at Frontier Machines - Beauty2019 -\\
		29 September / 4 October, 2019\\
		Ljubljana, Slovenia}

\begin{document}

\section{Introduction}

\begin{figure}[tb]
\centering
\begin{picture}(63,56)(-5,13)
   \put(0,50){\linethickness{3mm}\textcolor{red}{\line(1,0){10}}}
   \put(0,55){\linethickness{10mm}\textcolor{red}{\line(1,0){10}}}
   \put(3,53.5){\LARGE \textcolor{white}{$u$}}
   \put(11,55){\linethickness{10mm}\textcolor{blue}{\line(1,0){10}}}
   \put(11,50.5){\linethickness{1.05mm}\textcolor{red}{\line(1,0){9.5}}}
   \put(11,51.5){\linethickness{1.05mm}\textcolor{red}{\line(1,0){8.5}}}   
   \put(11,52.5){\linethickness{1.05mm}\textcolor{red}{\line(1,0){7.5}}}
   \put(11,53.5){\linethickness{1.05mm}\textcolor{red}{\line(1,0){6.5}}}   
   \put(11,54.5){\linethickness{1.05mm}\textcolor{red}{\line(1,0){5.5}}}
   \put(11,55.5){\linethickness{1.05mm}\textcolor{red}{\line(1,0){4.5}}}   
   \put(11,56.5){\linethickness{1.05mm}\textcolor{red}{\line(1,0){3.5}}}
   \put(11,57.5){\linethickness{1.05mm}\textcolor{red}{\line(1,0){2.5}}}   
   \put(11,58.5){\linethickness{1.05mm}\textcolor{red}{\line(1,0){1.5}}}
   \put(11,59.5){\linethickness{1.05mm}\textcolor{red}{\line(1,0){0.5}}}
   \put(15,53.5){\LARGE \textcolor{white}{$c$}}
   \put(22,55){\linethickness{10mm}\textcolor{pink}{\line(1,0){10}}}
   \put(26,53.5){\LARGE \textcolor{white}{$t$}}

   \put(34,55){\linethickness{10mm}\textcolor{red}{\line(1,0){10}}}
   \put(37,53.5){\LARGE \textcolor{white}{$g$}}
   \put(0,44){\linethickness{10mm}\textcolor{red}{\line(1,0){10}}}
   \put(3,42.5){\LARGE \textcolor{white}{$d$}}
   \put(3,48.5){\linethickness{4mm}\textcolor{white}{\line(1,1){5}}}
   \put(11,44){\linethickness{10mm}\textcolor{red}{\line(1,0){10}}}
   \put(15,42.5){\LARGE \textcolor{white}{$s$}}
   \put(22,44){\linethickness{10mm}\textcolor{blue}{\line(1,0){10}}}
   \put(25,42.5){\LARGE \textcolor{white}{$b$}}
   \put(34,44){\linethickness{10mm}\textcolor{lightgray}{\line(1,0){10}}}
   \put(37.5,42.5){\LARGE $\gamma$}
   \put(0,32){\linethickness{10mm}\textcolor{darkgreen}{\line(1,0){10}}}
   \put(3,30.5){\LARGE \textcolor{white}{$\nu_e$}}
   \put(11,32){\linethickness{10mm}\textcolor{darkgreen}{\line(1,0){10}}}
   \put(14,30.5){\LARGE \textcolor{white}{$\nu_\mu$}}
   \put(22,32){\linethickness{10mm}\textcolor{darkgreen}{\line(1,0){10}}}
   \put(25,30.5){\LARGE \textcolor{white}{$\nu_\tau$}}
   \put(34,32){\linethickness{10mm}\textcolor{pink}{\line(1,0){10}}}
   \put(36,30.5){\LARGE \textcolor{white}{$Z^0$}}
   \put(0,21){\linethickness{10mm}\textcolor{darkgreen}{\line(1,0){10}}}
   \put(3,19.5){\LARGE \textcolor{white}{$e$}}
   \put(11,21){\linethickness{10mm}\textcolor{darkgreen}{\line(1,0){10}}}
   \put(14,19.5){\LARGE \textcolor{white}{$\mu$}}
   \put(22,21){\linethickness{10mm}\textcolor{darkgreen}{\line(1,0){10}}}
   \put(25,19.5){\LARGE \textcolor{white}{$\tau$}}
   \put(34,21){\linethickness{10mm}\textcolor{pink}{\line(1,0){10}}}
   \put(34.5,19.5){\LARGE \textcolor{white}{$W^\pm$}}
   \put(46,55){\linethickness{10mm}\textcolor{YellowOrange}{\line(1,0){10}}}
   \put(49,53.5){\LARGE \textcolor{white}{$H$}}
      \put(-5,20){\rotatebox{90}{leptons}}
      \put(-5,44){\rotatebox{90}{quarks}}
      \put(9,66){generations}
      \put(3.2,62){I}
      \put(14.1,62){II}
      \put(25,62){III}
      \put(34.5,66){gauge}
      \put(34.5,62){forces}
      \put(46.5,66){Higgs}
      \put(46.5,62){boson}
\end{picture}
\caption{The SM in a typical lattice simulation focusing on calculations in $D$- and $B$-physics. Red indicates the quark flavors and gluons entering both sea and valence sector of the calculation, whereas heavy charm and bottom quarks (blue) may only be present in the valence sector. Top quarks and weak gauge bosons (magenta) enter the calculation as part of ``point like operators'' and leptons (green) contribute mostly in post-analysis steps. Effects of photons, iso-spin breaking, QED can be included but are mostly only required for sub-percent level precision. }
\label{fig.sketchSM}
\end{figure}
Lattice quantum chromodynamics (QCD) allows for first principle calculations of processes involving the strong force coupling quarks and gluons. Key features of lattice QCD calculations are that the method is also valid in the nonperturbative regime and systematical procedures exist to reduce uncertainties. As the name indicates, these calculations are based on discretizing the 4-dimensional space time to a rectangular grid, the lattice. After performing a Wick-rotation to Euclidean time, the QCD Lagrangian is simulated using the path integral formalism
\begin{align}
        \langle {\cal O} \rangle_E = \frac{1}{Z} \int {\cal D}[\psi,\overline{\psi}]\,{\cal D}[U]\, {\cal O}[\psi,{\overline{\psi}}, U]\, e^{-S_E[\psi,{\overline{\psi}},U]},
\end{align}
which results in a large but finite dimensional integral. The solution is obtained stochastically performing Monte Carlo simulations in a finite box of extent $\left(L/a\right)^3 \times T/a$, where $L$ is the number of lattice sites in the spatial $x$-, $y$-, $z$-direction and $T$ sites in the temporal $t$-direction. Any calculation is necessarily performed at a finite value of the lattice spacing $a$. On the one hand, a finite value of $a$ acts as an ultraviolet regulator and simulated quark masses $m_q$ in lattice units need to obey
\begin{align}
  am_q<1.
  \label{Eq.amq}
\end{align}
On the other hand, the finite spatial volume introduces an infrared regulator and we study physics in a finite box of size $L^3$. To keep e.g.~finite volume effects under control, an approximate measure is given by the simulated pion mass $aM_\pi$ times the spatial extent $L/a$. The empirically justified rule-of-thumb 
\begin{align}
  M_\pi \cdot L \lesssim 4,
  \label{Eq.MpiL}
\end{align}
suggests for a given spatial extent $L/a$ a minimum $aM_\pi$ such that finite volume effects are typically sufficiently small. A practical calculation requires further to choose between different discretizations of gauge and fermion actions like Wilson, Symanzik, or Iwasaki gauge actions and Wilson, Kogut-Susskind (staggered), domain wall, or overlap fermion actions.  Numerous  variations of these actions exist exhibiting different levels of improvements to reduce discretization or lattice artifacts. However, by performing simulations at different values of the lattice spacing $a$ and properly renormalizing all quantities, results at finite lattice spacing can be extrapolated to the continuum limit and are thus free of discretization effects. Continuum limit results are expected to agree regardless which discretization (lattice actions) have been used during the numerical simulation and can hence be meaningfully compared to experimental results or other phenomenological calculations. Further details on the techniques how to perform numerical lattice simulations or extract physical continuum limit results are explained in various text books, see e.g.~\cite{Montvay:1994cy,DeGrand:2006zz,Luscher:2010ae,Knechtli:2017sna}.

Most state-of-the-art lattice QCD simulations are based on a set of gauge field configurations containing the effects of dynamical 2+1 or 2+1+1 flavors in the sea-sector. This means two degenerate up/down quarks, strange and eventually charm quarks contribute in the sea-sector and are part of the evolution algorithm generating the set of gauge field configurations. These set of gauge field ensembles typically include some ensembles featuring a physical value of the pion mass whereas strange and, if present,  charm are always close to their physical value. Subsequently these gauge field configurations are then used to perform valence sector measurements where in addition to light, strange and charm quarks also bottom quarks are simulated. Due to their large mass, simulations with charm and bottom quarks face an additional challenge and depending on the lattice spacing may require to use an effective action (e.g.~non-relativistic QCD, Fermilab or relativistic heavy quark action) to avoid the bound in Eq.~(\ref{Eq.amq}). Alternatively, simulations below the physical value of the bottom mass are performed combined with a (benign) extrapolation (e.g.~heavy HISQ or heavy domain wall formalism).

In Fig.~\ref{fig.sketchSM} we sketch the SM from the point of view of a lattice QCD simulation showing the dynamical quarks and gluons in red and the heavy flavors present only in the valence sector in blue. Top quark and electroweak gauge bosons (magenta) typically enter in point-like operators implementing e.g.~short distance contributions of the weak force, whereas leptons enter the calculations as part of the post-simulation analysis. These are the most relevant ingredients for the calculations subject to this overview which focuses at certain heavy-light calculations i.e.~calculations with a charm or bottom valence quark. Effects due to quantum electrodynamics (QED), electric charges, or photons are typically small and become only relevant at the sub-percent level of precision not yet reached for the heavy-light calculations presented here. In the following we will therefore simplify the notation and suppress e.g.~electric charges. Interactions with the Higgs boson can be calculated as additional operators to determine e.g.~contriubtions of theories describing physics beyond the SM but are not subject of this review.

As mentioned above, lattice determinations extrapolated to the continuum limit are expected to mutually agree and similar to several experimental measurements a meaningful average can be defined. For a large set of phenomenologically relevant quantities these averages are provided by the Flavor Lattice Averaging Group (FLAG)\footnote{\href{http://flag.unibe.ch}{http://flag.unibe.ch}} which scrutinizes the calculations to meet specified quality criteria and calculates averages accounting e.g.~if any two calculations are (partly) correlated. In addition the different methods used for the determination of the quantities are reviewed.

In the following we will first in Section \ref{Sec.psFF} report on semi-leptonic charm and beauty form factors with a single, pseudoscalar hadronic final state before discussing beauty form factors with a single vector final state and the determination of $R(D^*)$ in Section \ref{Sec.vtFF}. Subsequently we briefly summarize the status of the determination of bottom and charm quark masses in Section \ref{Sec.c_b_mass} and report updates on neutral $B_{(s)}$ meson mixing in Section \ref{Sec.mixing}. We end with a summary also highlighting further new developments.

\section{Charm and beauty form factors}\label{Sec.psFF}
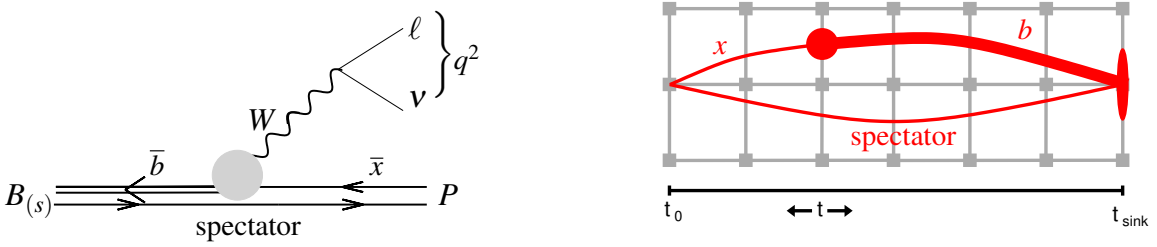
\begin{figure}[tb]
  \begin{picture}(66,30)
    \put(6,2){\includegraphics[width=50mm]{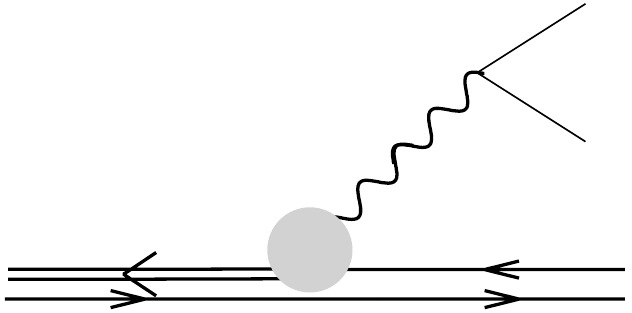}}
    \put(0,3){\large{$B_{(s)}$}} \put(57,3){\large{$P$}}
    \put(32,13){\large{$W$}} 
    \put(53,25){\large{$\ell$}} \put(53,16){\large{$\nu$}}
    \put(56,22){$\Bigg\}$}\put(59,21){$q^2$}
    \put(25,-1){spectator}
    \put(48,7){$\overline{x}$}
    \put(19,7){$\overline{b}$}
  \end{picture}
  \hfill
  \begin{picture}(65,30)
    \put(0,0){\includegraphics[width=65mm]{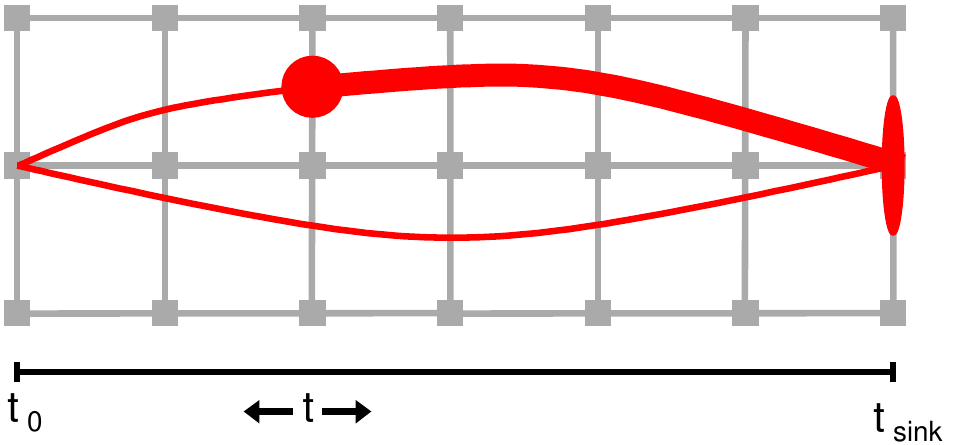}}
    \put(47,25){\textcolor{red}{$b$}}
    \put(7,23){\textcolor{red}{$x$}}
    \put(25,11){\textcolor{red}{\textrm{spectator}}}
  \end{picture}
  \caption{Left: Sketch of tree-level weak semi-leptonic $B_{(s)}$ decays
    mediated by a charged $W$ boson in a setup with the $B_{(s)}$
    meson at rest. $P$ denotes the pseudoscalar final state ($\pi$, $K$,
    $D$, or $D_s$), the spectator is a light up/down or a strange
    quark, and $x$ the up or charm daughter quark. The momentum transferred to the leptons is given by $q^2= M_{B_{(s)}}^2 + M_P^2 - 2 M_{B_{(s)}} E_P$.
    Right: Schematic lattice calculation of the corresponding hadronic weak decay matrix element.}
  \label{fig.treelevel}
\end{figure}

Conventionally semi-leptonic decays with a single pseudoscalar hadronic final state are para\-me\-trized by relating the experimentally measured branching fraction 
\begin{multline}
  \frac{d\Gamma(B_{(s)}\to P\ell\nu)}{dq^2} = \frac{\eta_{EW} G_F^2 |V_{xb}|^2}{24\pi^3}\frac{(q^2-m_\ell^2)^2\sqrt{E^2_P - M^2_P}}{
q^4 M_{B_{(s)}}^2} \\
  \times \Bigg[ \bigg(\!1 + \frac{m^2_\ell}{2q^2}\! \bigg)M^2_{B_{(s)}}(E_P^2 - M_P^2) |f_+(q^2)|^2 + 
    \frac{3m^2_\ell}{8q^2}(M_{B_{(s)}}^2 - M_P^2)^2 |f_0(q^2)|^2  \Bigg]\,,
  \label{Eq.BF}
\end{multline}
to two form factors, $f_+$ and $f_0$, the nonperturbative contribution of the strong force. In Eq.~(\ref{Eq.BF}) we show the general expression for a pseudoscalar $B_{(s)}$ meson decaying at rest to a pseudoscalar meson $P$ and a lepton-neutrino pair. This process sketched on the left in Fig.~\ref{fig.treelevel} is a tree-level weak decay in the SM in which a $b$ quark decays to an up or charm quark (generally referred to as $x$) under the emission of a $W$ boson. Hence the Cabibbo-Kobayashi-Maskawa (CKM) matrix element $|V_{xb}|$ enters in Eq.~(\ref{Eq.BF}) in addition to the mass of the $B_{(s)}$ meson, $M_{B_{(s)}}$ as well as the energy and mass of the final state hadron, $E_P$ and $M_P$.  The mass of the final state lepton $\ell$  with $\ell=e,\,\mu,\,\tau$ is denoted by $m_\ell$ and the lepton-neutrino pair acquires the 4-momentum $q^\mu$.  

To determine the form factors $f_+$ and $f_0$, we evaluate the hadronic matrix element
\begin{align}
\langle P(p_P) | {\cal V}^\mu |B_{(s)}(p_{B_{(s)}})\rangle = f_+(q^2)\left( p^\mu_{B_{(s)}} + p^\mu_P - \frac{M_{B_{(s)}}^2 - M_P^2}{q^2}q^\mu \right) + f_0(q^2)\frac{M_{B_{(s)}}^2 - M_P^2}{q^2}q^\mu\,,
\end{align}
where $p_{B_{(s)}}$ and $p_P$ are the momenta of the $B_{(s)}$ and $P$ meson, respectively, and ${\cal V}^\mu = \bar x \gamma^\mu b$ is the weak decay operator obtained after performing an operator product expansion (OPE) to identify the short distance contribution. Implementing the hadronic weak decay matrix element on the lattice, we extract the form factors by measuring 3-point functions. As schematically shown by the quark line diagram on the right in Fig.~\ref{fig.treelevel}, one possibility to implement the calculation is to consider the initial $B_{(s)}$ meson at $t_\text{sink}$ and the final state pseudoscalar at $t_0$. The spectator quark remains unchanged, whereas the $b$ quark changes into an $x$ quark at time $t$ with $t_0\le t \le t_\text{sink}$.\footnote{Placing the $B_{(s)}$ meson at $t_\text{sink}$ reduces in this setup the numerical costs because e.g.~$B\to\pi\ell\nu$ decays can be calculated with one cheap $b$ quark and one expensive $u/d$ quark inversion used for both spectator and daughter quark.}  The different steps to obtain form factors over the full kinematically allowed $q^2$ range will be outlined for $B\to\pi\ell\nu$ decays and we subsequently summarize the status for other decays and highlight recent developments.

\subsection{$B\to \pi\ell\nu$}
\begin{figure}[tb]
  \includegraphics[width=0.49\textwidth]{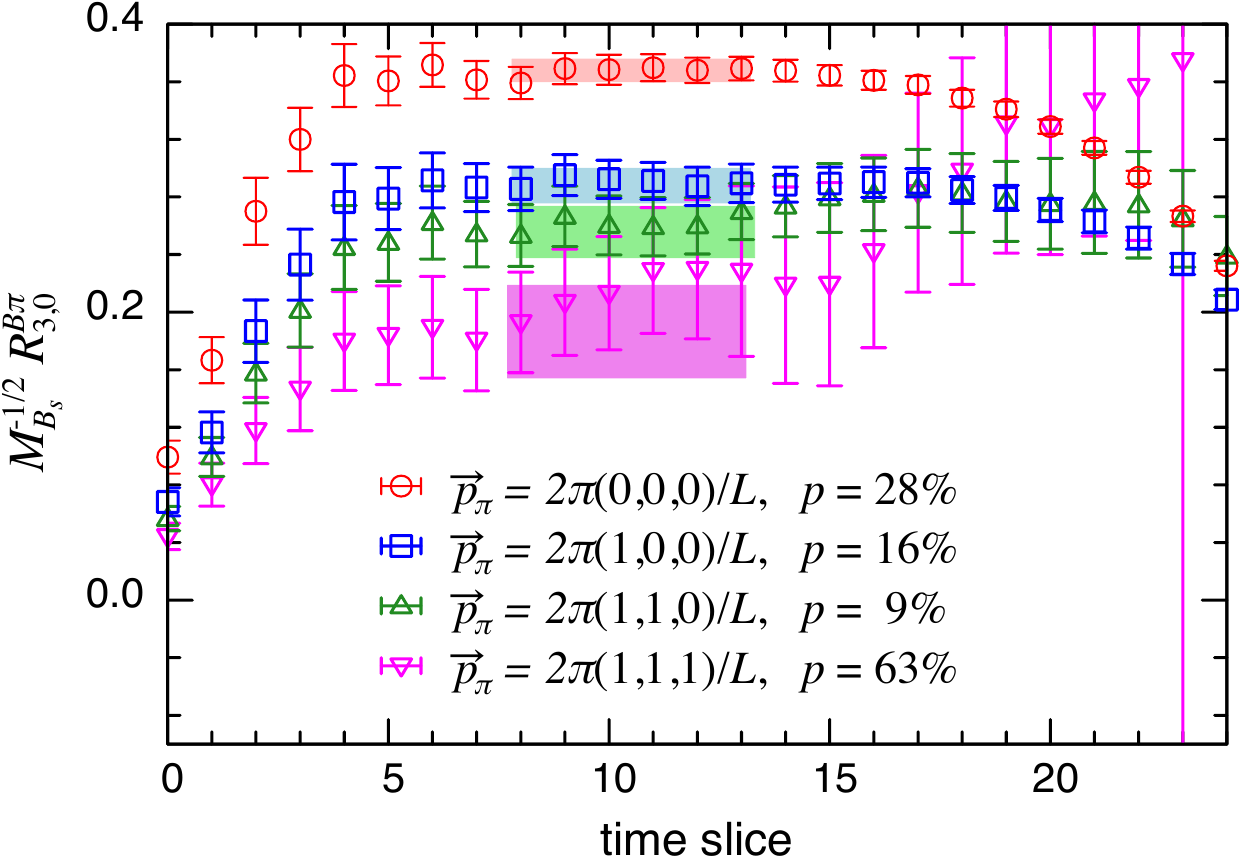}
  \hfill
  \includegraphics[width=0.49\textwidth]{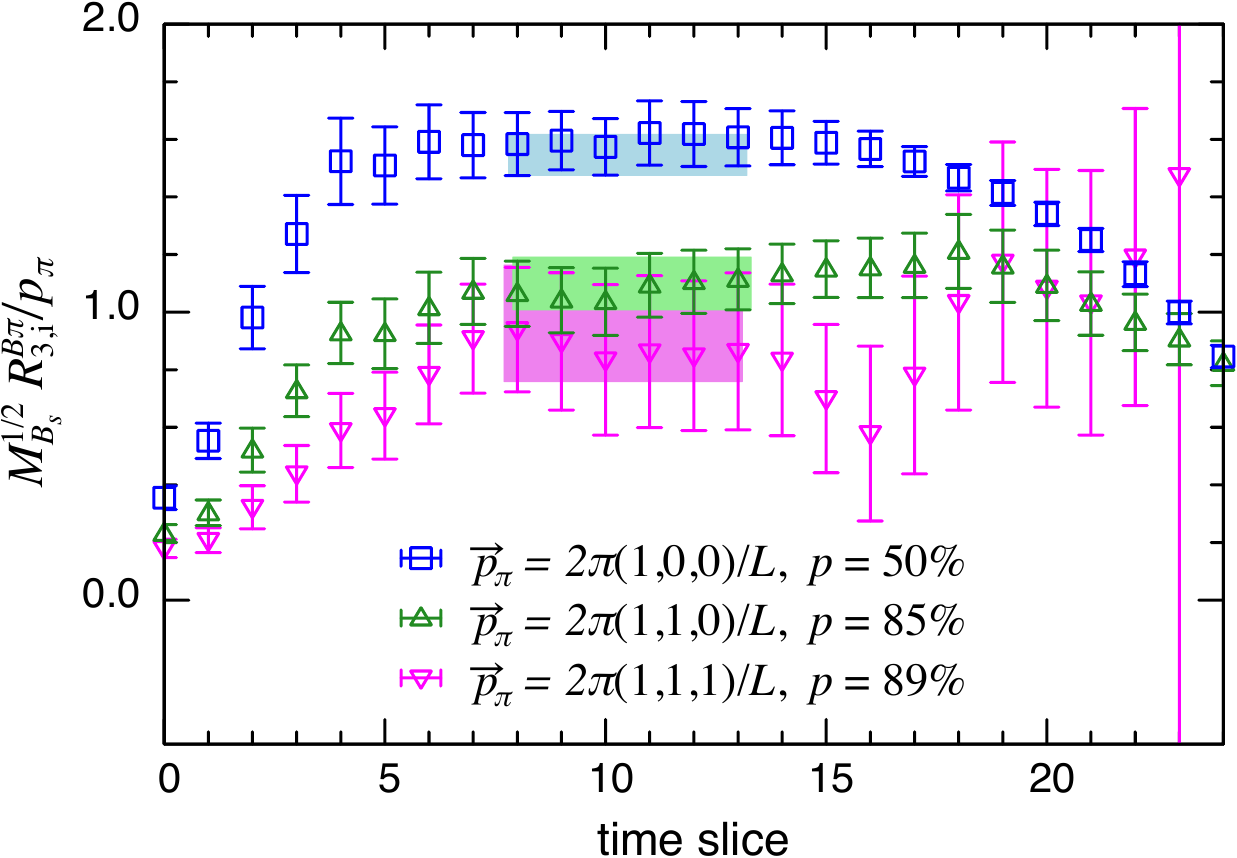}
  \caption{Example from Ref.~\cite{Flynn:2015mha} to extract values for $B\to\pi\ell\nu$ form factors on one ensemble by performing a fit to the ground state in the indicated plateau range. The different colors show the projection onto final state pions with different units of discretized momentum.}
  \label{fig.Bpi_ratio}
\end{figure}

The calculation of $B\to\pi\ell\nu$ form factors proceeds along the sketches in Fig.~\ref{fig.treelevel} by using a propagator with $u/d$ quark mass for the spectator as well as the daughter quark. As mentioned above, we need to perform calculations using ensembles of gauge field configurations at different values of the lattice spacing to perform a controlled continuum limit extrapolation. In addition it is helpful to have ensembles also varying the value of the degenerate $u/d$ quark mass to guide, if needed, the chiral extrapolation. For each of these ensembles, the aforementioned 3-point functions need to be calculated and in addition 2-point functions describing the hadronic initial and final state meson.  Appropriate ratios of 3-point over 2-point functions allow then to extract form factors. An example for this determination is shown in Fig.~\ref{fig.Bpi_ratio}. In this example only the ground state signal is extracted from a fit to the indicated plateau region. However, also excited state contributions related to the pion on the left and/or the $B$ meson on the right can be included in the determination. 

\begin{figure}[tb]
  \includegraphics[width=0.49\textwidth]{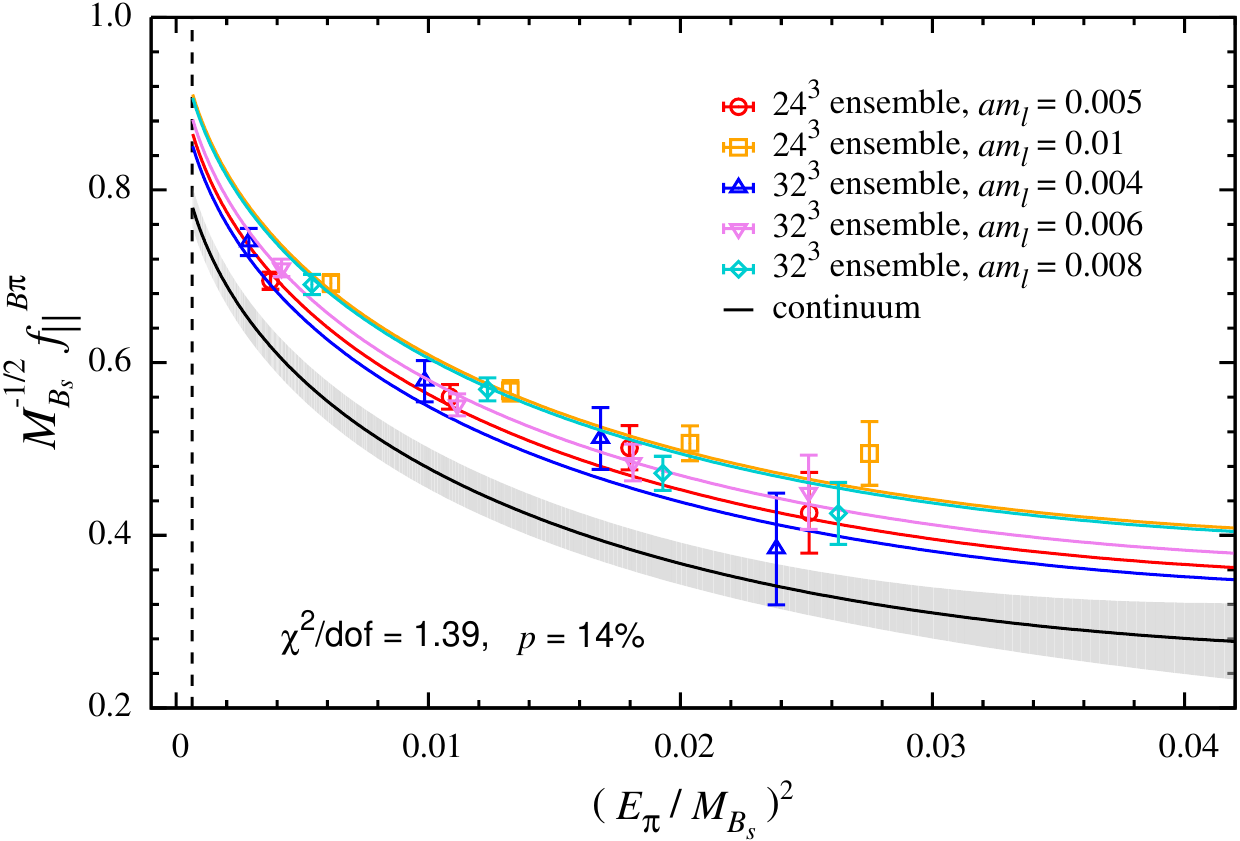}
  \hfill
  \includegraphics[width=0.49\textwidth]{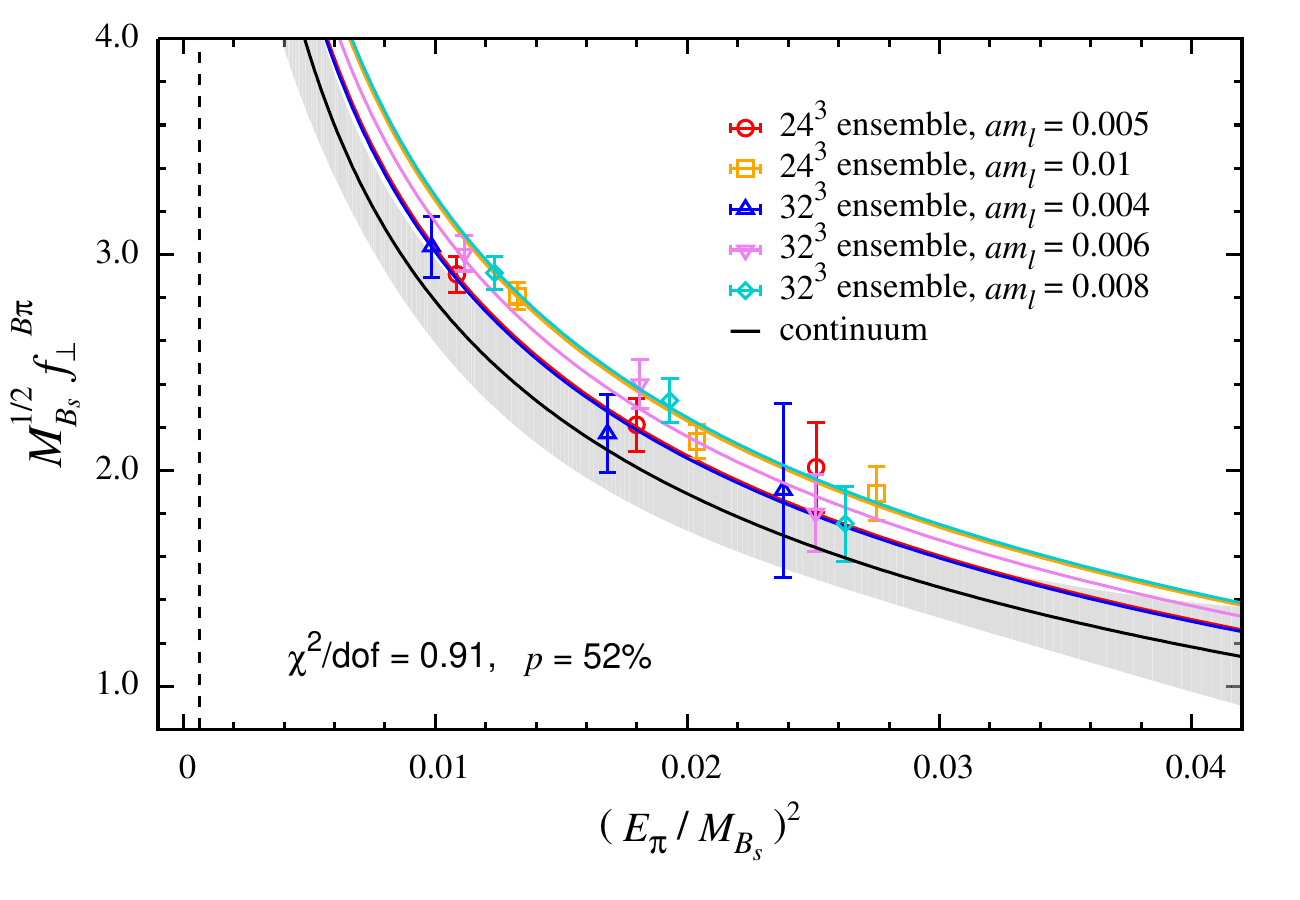}
  \caption{Example of obtaining the chiral- and continuum limit (black line with gray error band) for by performing a fit to the renormalized $B\to \pi\ell\nu$ form factors. The fit combines the data from five different gauge field ensembles (colored data points) and the colored lines indicate the fit result corresponding to the parameters of the ensemble. Plots taken from from Ref.~\cite{Flynn:2015mha}.}
  \label{fig.Bpi_ChiPTfit}
\end{figure}

\begin{figure}[tb]
  \includegraphics[width=0.49\textwidth]{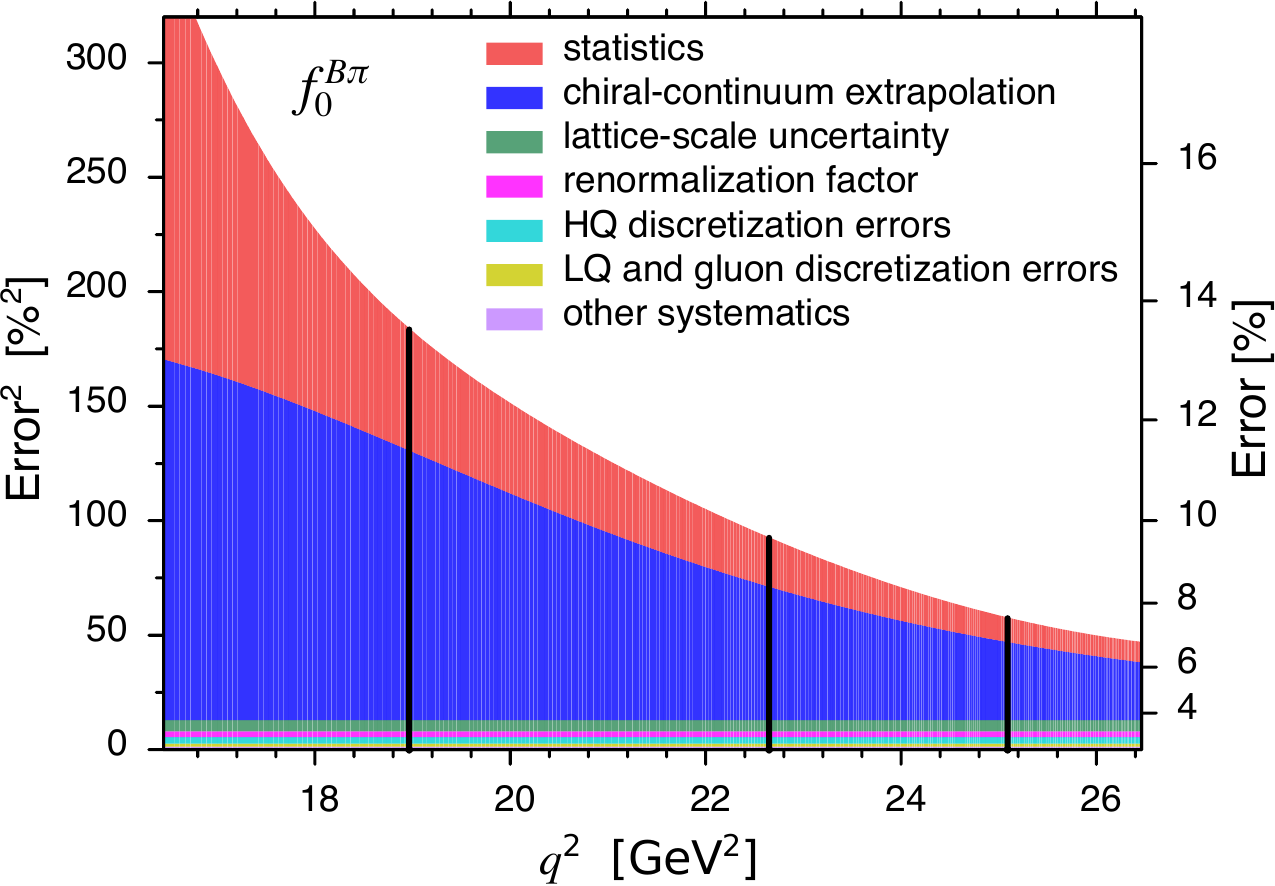}
  \hfill
  \includegraphics[width=0.49\textwidth]{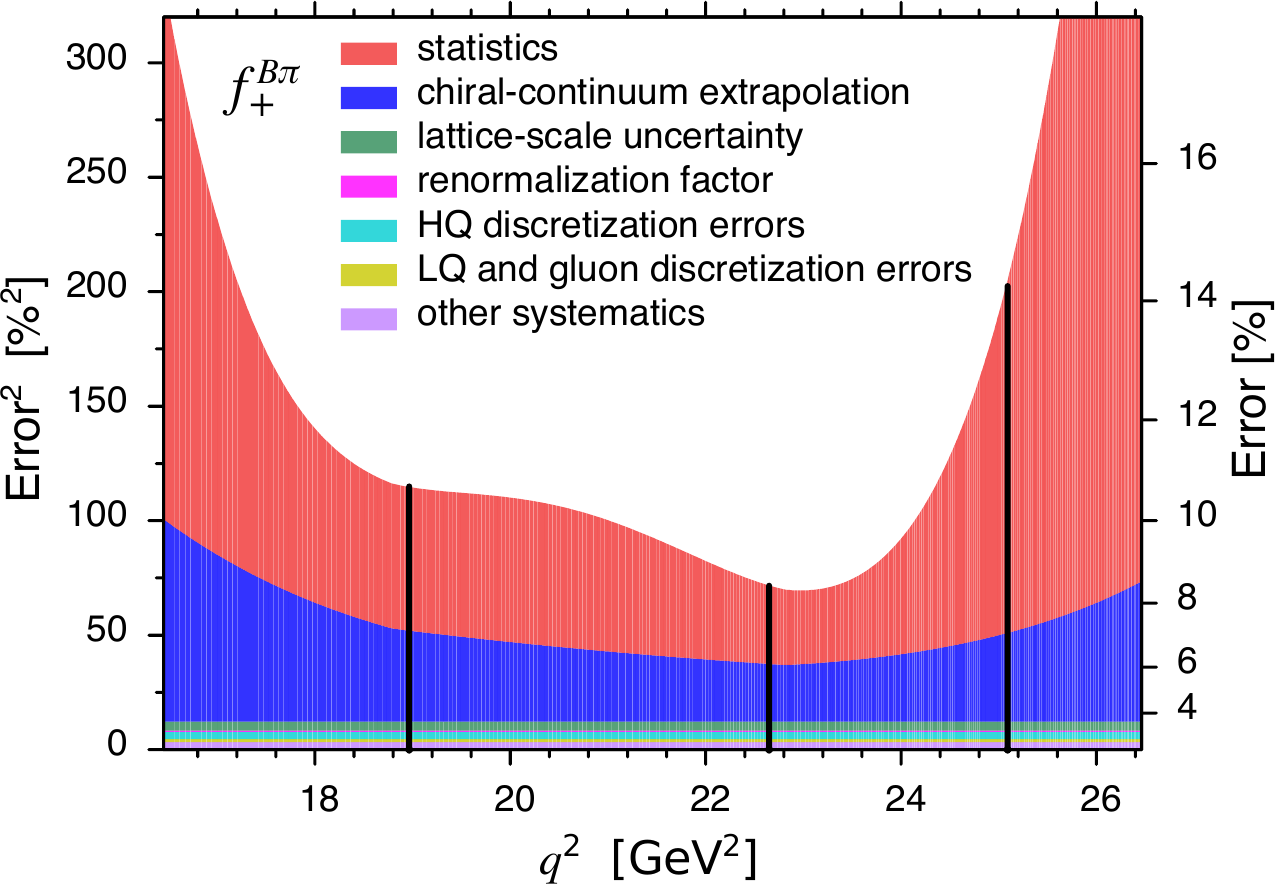}
  \caption{Graphical error budget showing statistical and systematic uncertainties for the calculation of $B\to\pi\ell\nu$ form factors from Ref.~\ref{fig.Bpi_error}. The vertical black lines indicate values of the ``synthetic data points'' used in the subsequent $z$-expansion.}
  \label{fig.Bpi_error}
\end{figure}

After extracting form factors for all ensembles, corresponding renormalization factors need to be multiplied before combining the data from different gauge field ensembles in the next step. In Fig.~\ref{fig.Bpi_ChiPTfit} the renormalized form factors obtained for five different gauge field ensembles at two different values of the lattice spacing are shown. Red and yellow data correspond to the coarser ensemble, blue, magenta and cyan to the ensembles with finer lattice spacing. Using a fit ansatz inspired by heavy meson chiral perturbation theory in the hard pion limit, the extrapolation to the continuum and physical quark masses is performed and shown by the black line with gray error band.
So far only the statistical uncertainties directly originating from the numerical simulation have been accounted for. Further systematic effects need to be considered. A graphical error budget showing the different uncertainties contributing to the calculation of Ref.~\cite{Flynn:2015mha} is presented in Fig.~\ref{fig.Bpi_error}. These plots also show the range in $q^2$ covered by the numerical lattice simulation. By construction, lattice simulations prefer the zero recoil limit where the final state hadron is at rest and most of the released energy of the $B$ meson is transferred to the lepton pair. Within the covered range of $q^2$, vertical black lines indicate three $q^2$ values which enter as so called ``synthetic data points'' in the final step of the form factor analysis, the $z$-expansion to obtain form factors over the full $q^2$ range.\footnote{Some authors prefer a different strategy and perform a single chiral-continuum-kinematical extrapolation.} 

\begin{figure}[tb]  
  \includegraphics[width=0.49\textwidth]{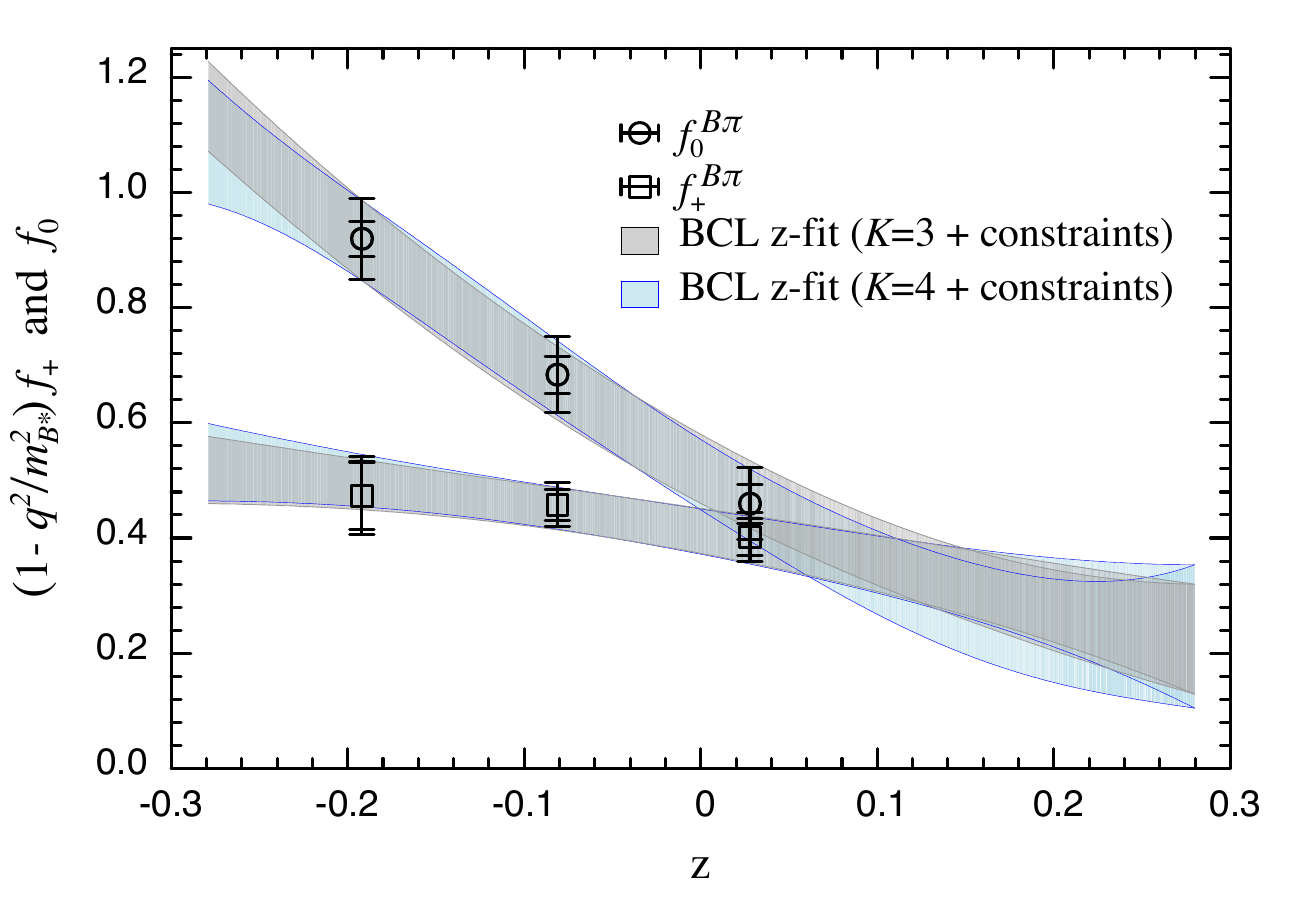}
  \hfill
  \includegraphics[width=0.49\textwidth]{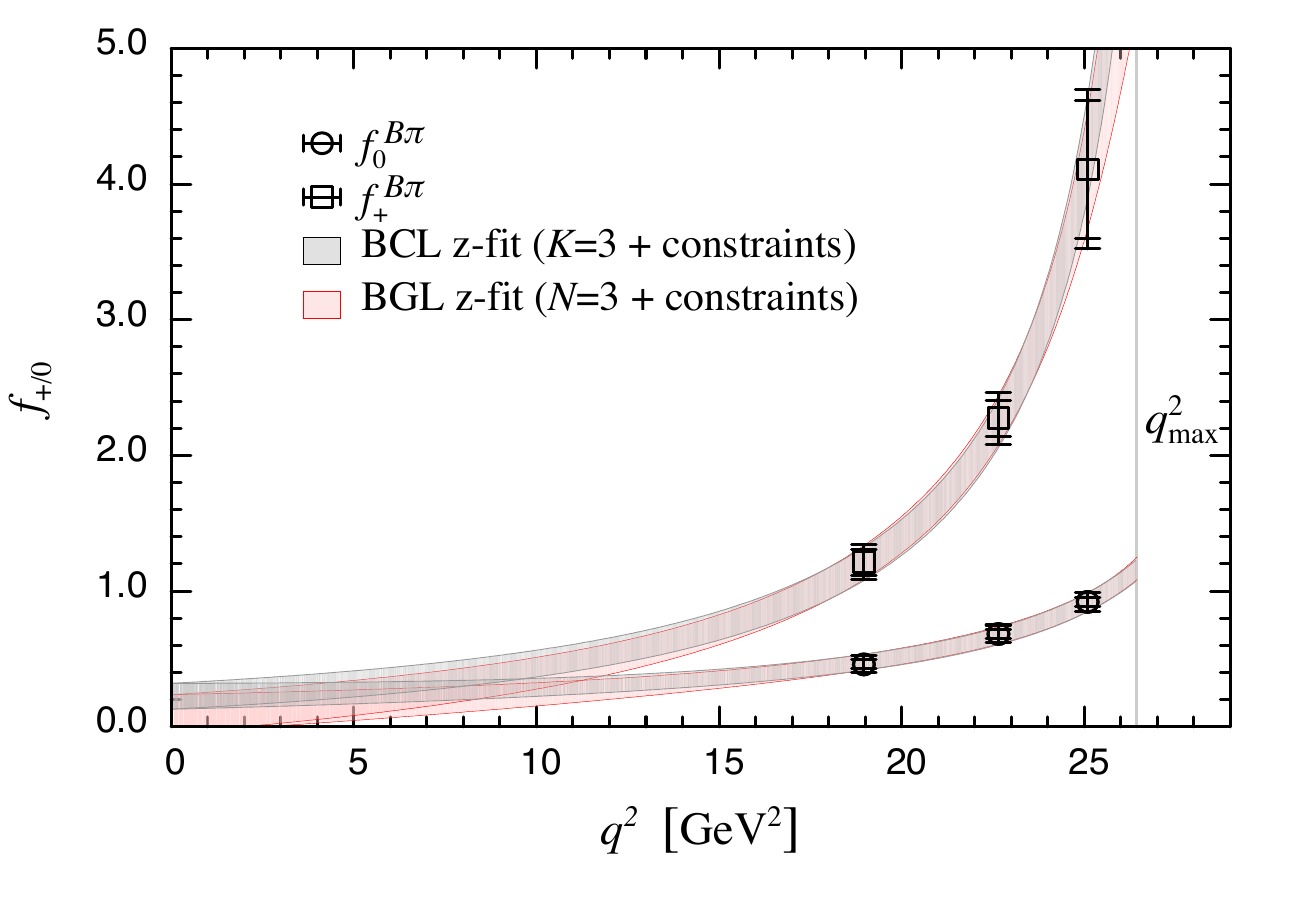}
  \caption{Kinematical extrapolation over the full $q^2$ range performed in terms of a BCL $z$-expansion on three synthetic data points from Ref.~\cite{Flynn:2015mha}. The plot on the left shows the extrapolation in $z$ space; the plot on the right is converted back to $q^2$.}
  \label{fig.Bpi_zfit}
\end{figure}

\begin{figure}[tb]  
  \includegraphics[height=0.2\textheight]{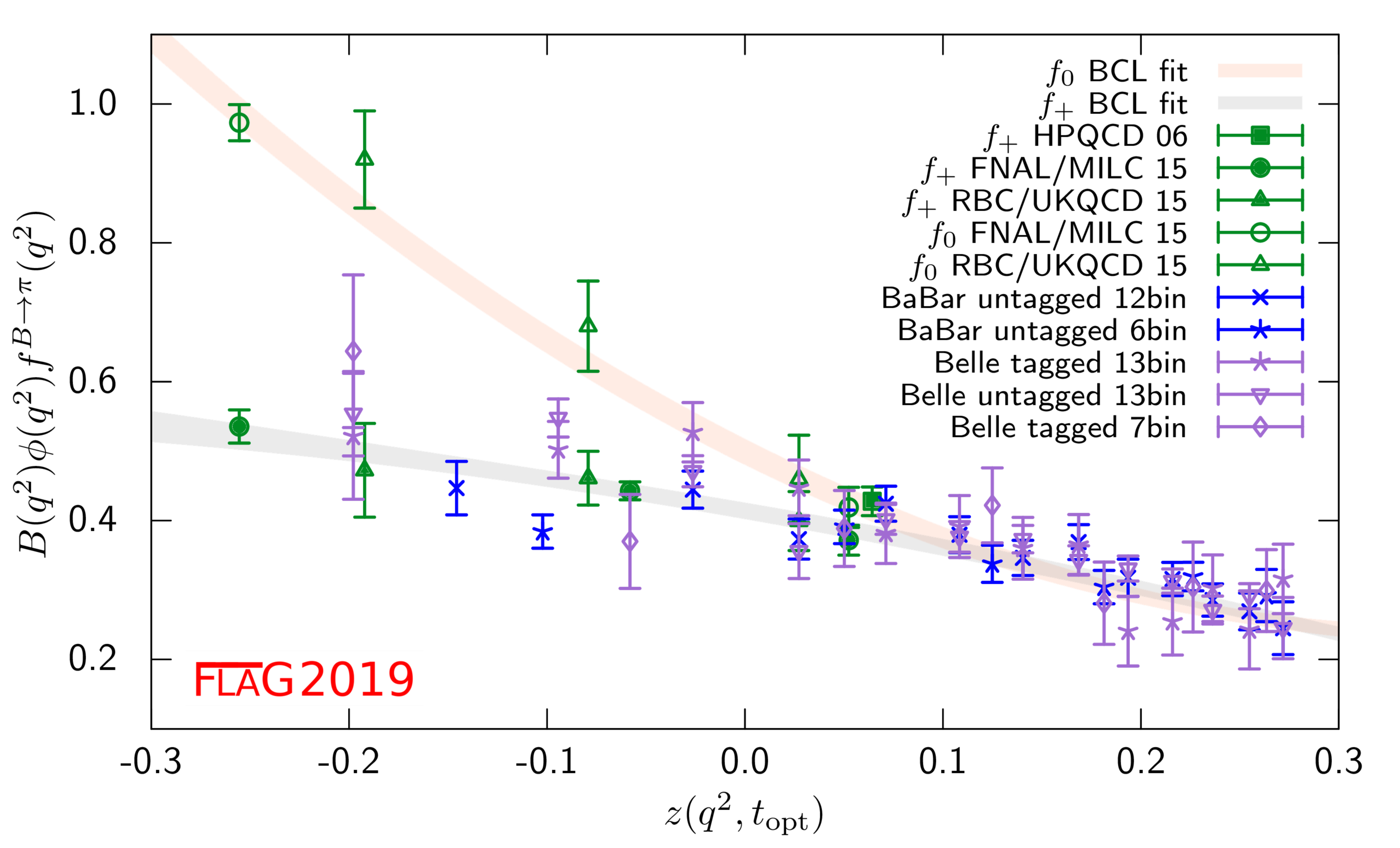}
  \hfill
  \includegraphics[height=0.2\textheight]{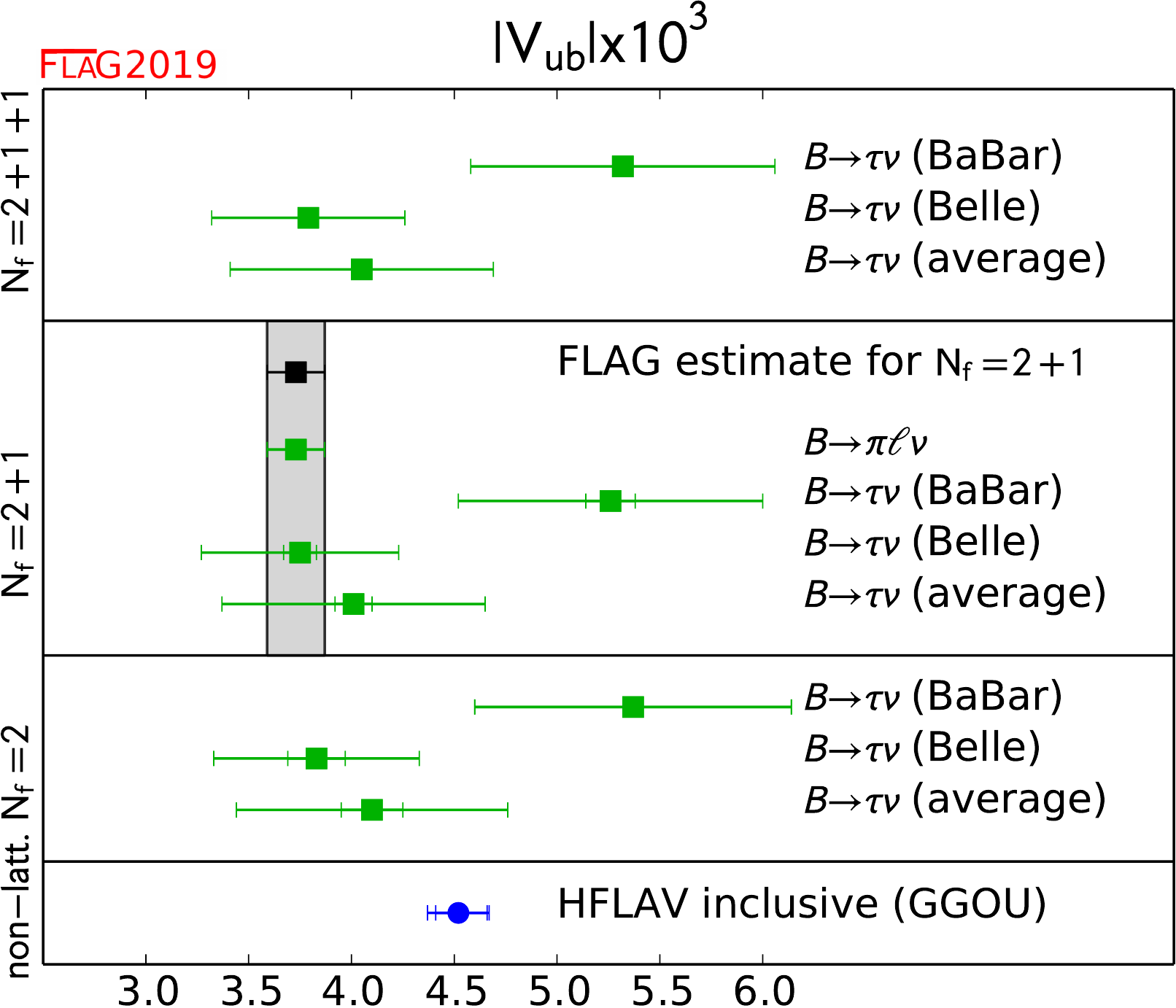}
  \caption{Combined FLAG 2019 analysis \cite{Aoki:2019cca} to determine the CKM matrix element $|V_{ub}|$ by performing a combined analysis to three lattice determinations of $B\to\pi\ell\nu$ form factors \cite{Flynn:2015mha,Dalgic:2006dt,Lattice:2015tia} and the BaBar \cite{delAmoSanchez:2010af,Lees:2012vv} and Belle \cite{Ha:2010rf,Sibidanov:2013rkk} experimental data sets.}
  \label{fig.Bpi_flag}
\end{figure}

In the case of $B\to\pi\ell\nu$ semi-leptonic decays commonly the $z$ expansions by Boyd, Grinstein, Lebed (BGL) \cite{Boyd:1994tt} and Bourrely, Caprini, Lellouch (BCL) \cite{Bourrely:2008za} are considered. The general idea of $z$ parametrizations is to map the kinematically allowed range in $q^2$ to the unit disk and then perform an expansion in the new $z$ variable. As an example we show a kinematical extrapolation using the BCL parametrization for the three synthetic data points introduced above in Fig.~\ref{fig.Bpi_zfit}. The synthetic data points enter the $z$-expansion with combined statistical and systematic uncertainties. Hence the resulting form factors covering the full $q^2$ range can be meaningfully compared to other lattice determinations, results obtained e.g.~from QCD sum rules, or combined with experimental results to extract the CKM matrix element $|V_{ub}|$.
An example from the FLAG analysis \cite{Aoki:2019cca} based on three different lattice determinations \cite{Flynn:2015mha,Dalgic:2006dt,Lattice:2015tia} and the experimental data sets from BaBar \cite{delAmoSanchez:2010af,Lees:2012vv} and Belle \cite{Ha:2010rf,Sibidanov:2013rkk} is shown in Fig.~\ref{fig.Bpi_flag}. By performing a combined fit, the relative normalization between lattice and experimental data is obtained which is given by the CKM matrix element $|V_{ub}|$. The plot on the left demonstrates that for $B\to\pi\ell\nu$ only the combination of form factors obtained on the lattice and measured in experiments covers the entire range in $q^2$. Nevertheless the form factors of the combined fit are in excellent agreement with most of the data points. This fit determines
\begin{align}
  |V_{ub}^\text{excl}|=3.73(14)\times 10^{-3},
\end{align}
  which as shown on the right in Fig.~\ref{fig.Bpi_flag} is in perfect agreement with other exclusive determinations of $|V_{ub}|$ but exhibits a $2-3\sigma$ tension with the inclusive determination.

Several new and improved lattice calculations required to update the value of $|V_{ub}^\text{excl}|$ are currently pursued and preliminary results have already been reported at past Lattice conferences \cite{Colquhoun:2019tyq,Flynn:2019jbg,Gelzer:2019zwx,Lubicz:2018scy}.

\subsection{$B_s\to K\ell\nu$}
\begin{figure}[tb]
  \centering
  \includegraphics[height=0.2\textheight]{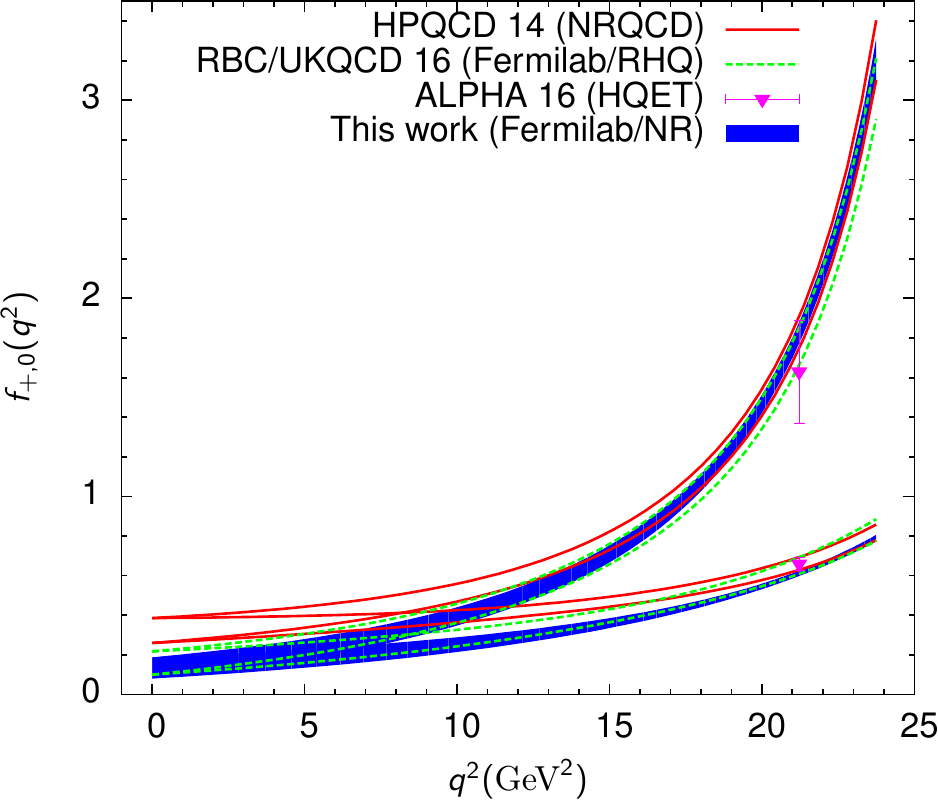}
  \hspace{20mm}
   \includegraphics[height=0.2\textheight]{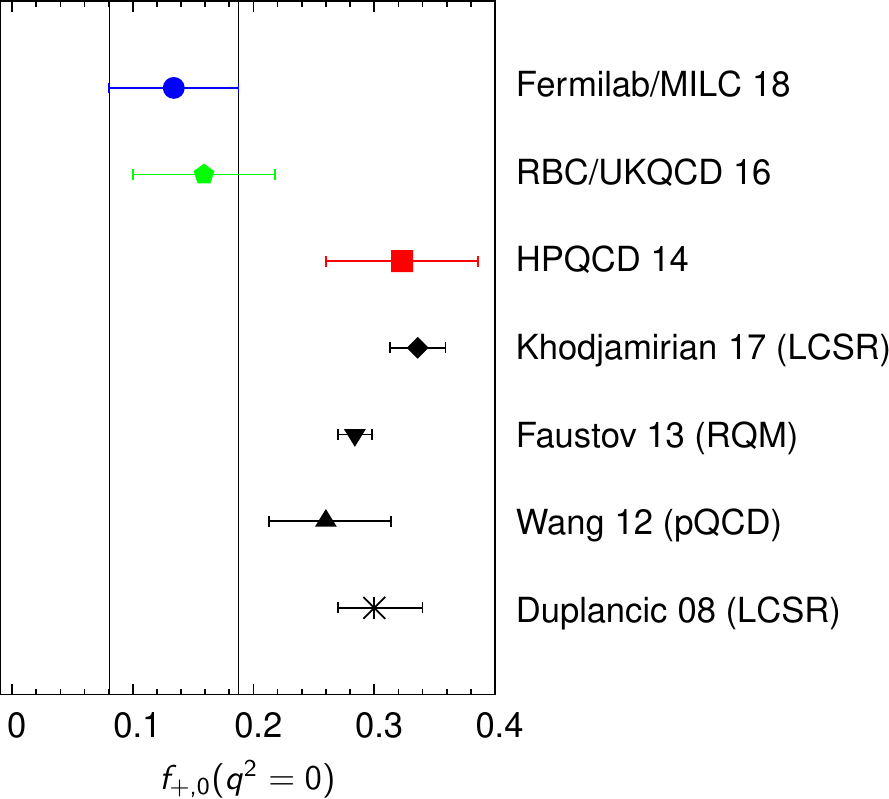} 
  \caption{Left: Comparison of $B_s\to K\ell\nu$ form factors as function of $q^2$ determined by different lattice collaborations \cite{Flynn:2015mha,Bouchard:2014ypa,Bazavov:2019aom,Bahr:2016ayy}. Right: Results at $q^2=0$ obtained either from a kinematical extrapolation of lattice results or form calculations based on analytic methods \cite{Duplancic:2008tk,Khodjamirian:2017fxg,Wang:2012ab,Faustov:2013ima}. Plots extracted from \cite{Bazavov:2019aom}.}
  \label{fig.BsK}
\end{figure}

If we change compared to the $B\to\pi\ell\nu$ form factor calculation only the spectator to a strange quark in the diagrams sketched in Fig.~\ref{fig.treelevel}, we determine form factors for $B_s\to K \ell \nu$ decays which also feature a bottom quark weakly decaying into an up quark. The lattice favors this channel to determine $|V_{ub}|$ because statistical noise grows when simulating quarks with lighter mass. For the same number of measurements, a kaon is hence more precise than a pion. Unfortunately, no experimental measurements for this channel have been published so far because the $B$ factories BaBar and Belle predominantly ran at the $\Upsilon(4s)$ threshold generating $B$ but not $B_s$ mesons. Fortunately, this limitation is not present at the Large Hadron Collider (LHC) and we look forward to a measurement by LHCb and hopefully also by Atlas or CMS. On the lattice three collaborations, HPQCD, RBC-UKQCD, and Fermilab/MILC, have published form factors extending over the full $q^2$ range \cite{Flynn:2015mha,Bouchard:2014ypa,Bazavov:2019aom} and in addition the Alpha collaboration has published a determination at one specific $q^2$ value \cite{Bahr:2016ayy}. A comparison of the lattice determinations performed by Fermilab/MILC is shown in the left plot of Fig.~\ref{fig.BsK}. In the plot on the right, the extrapolated lattice determinations at $q^2=0$ are compared to results obtained from light cone sum rules \cite{Duplancic:2008tk,Khodjamirian:2017fxg}, perturbative QCD \cite{Wang:2012ab} and relativistic quark model \cite{Faustov:2013ima} calculations. Some tension is present for the values at $q^2=0$ which desires a better understanding. Eventually updates of presently preliminary results \cite{Flynn:2019jbg,Gelzer:2019zwx} may help to address this.

\subsection{$B_s\to D_s\ell\nu$}
\begin{figure}[tb]
  \centering
   \includegraphics[height=0.25\textheight]{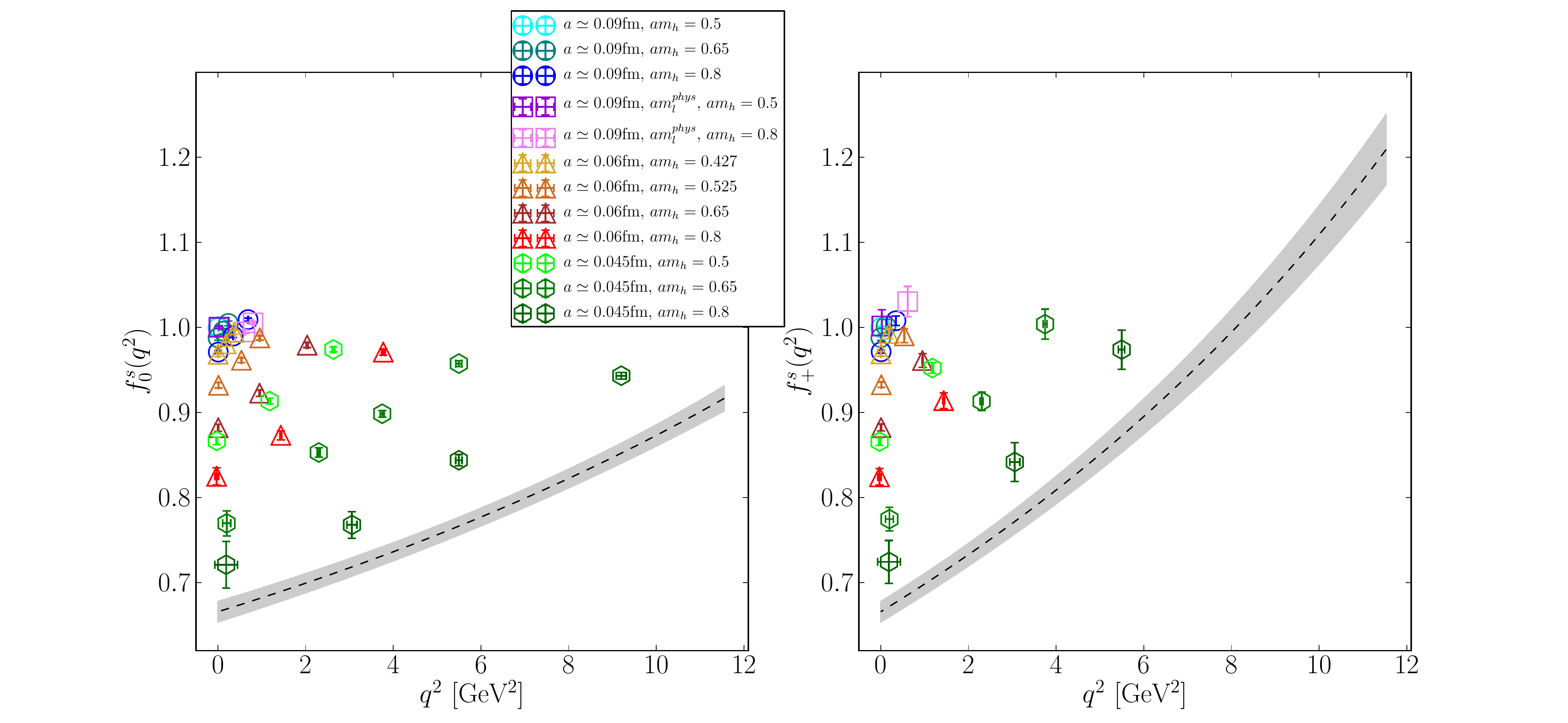}\\
   \includegraphics[height=0.19\textheight]{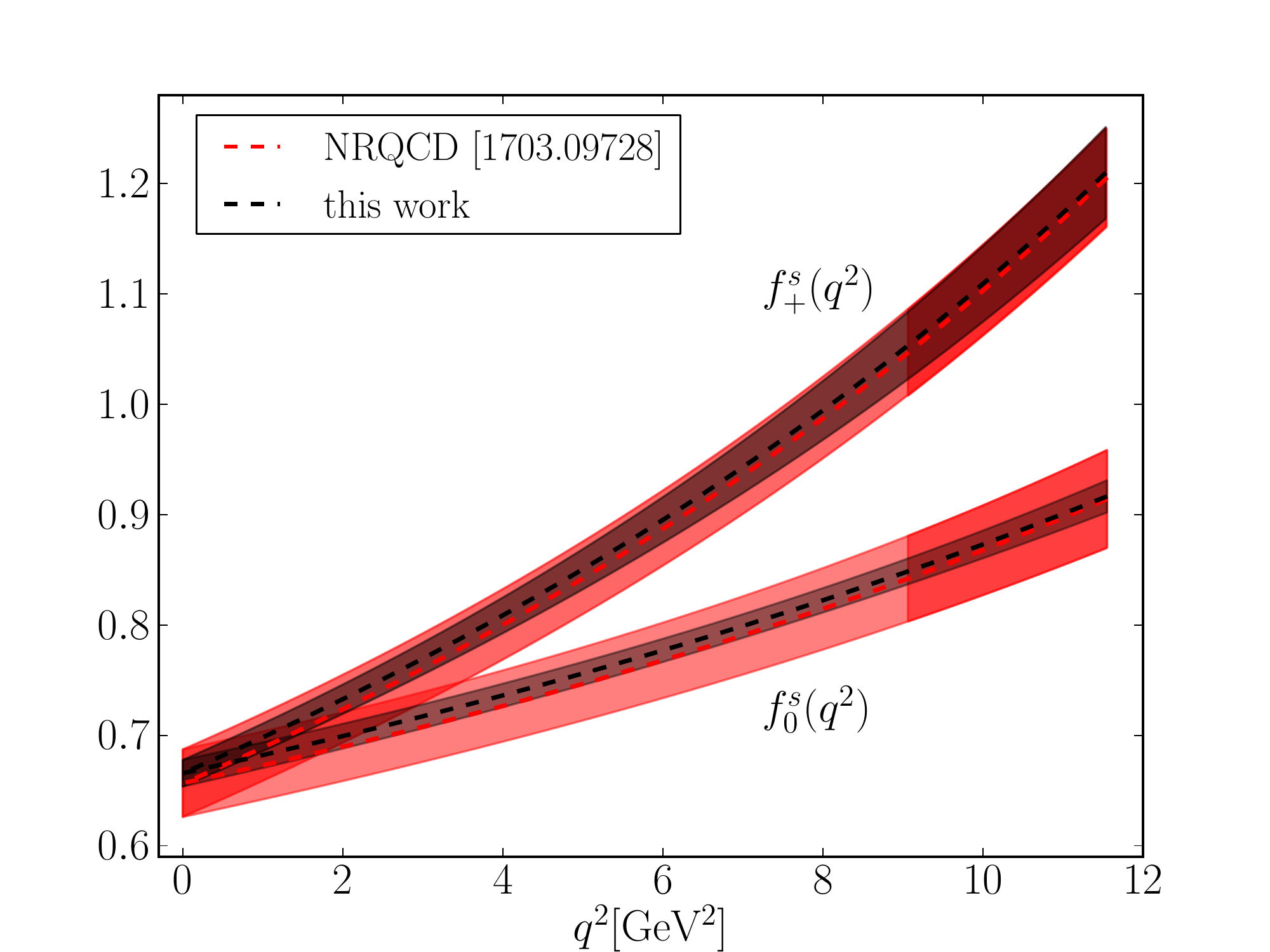}
   \hspace{10mm}
   \includegraphics[width=0.42\textwidth]{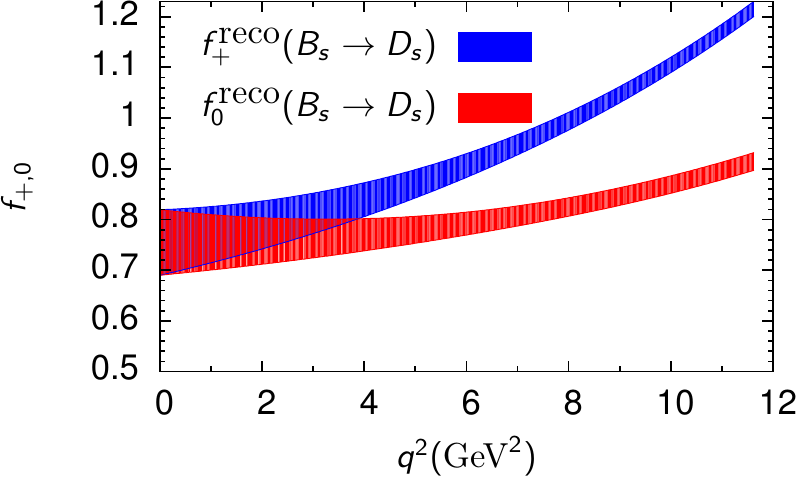}
   \caption{$B_s\to D_s\ell\nu$ semi-leptonic form factors. The upper plots demonstrate HPQCD's progress using the same action for all quarks and simulating $b$ quarks almost at their physical mass. Combined with simulations featuring an array of mass values between charm and bottom, it is possible to cover almost the entire range of $q^2$ in numerical lattice simulations at the finest lattice spacing of $a\simeq 0.045$ fm \cite{McLean:2019qcx}. The lower left plot shows how the new method reduces the uncertainties compared to HPQCD's previous determination \cite{Monahan:2017uby}; whereas the right plot presents the updated determination by the Fermilab/MILC collaboration \cite{Bazavov:2019aom,Bailey:2012rr}. Plots extracted from Refs.~\cite{McLean:2019qcx} and \cite{Bazavov:2019aom}.}
  \label{fig.BsDs}
\end{figure}

Replacing next the daughter quark in the diagrams in Fig.~\ref{fig.treelevel} with a charm quark, we can turn the calculation into the determination of $B_s\to D_s\ell\nu$ form factors. In these decays the $b$ quark weakly decays to a charm quark and in light of the recent measurement by LHCb \cite{Aaij:2020hsi} we explore a channel to determine the CKM matrix element $|V_{cb}|$. Since this channel features no light $u/d$ quarks in the valence sector, it is even more favorable for a lattice calculation. In fact, this channel is used by the HPQCD collaboration to demonstrate what conceptual progress in semi-leptonic form factor calculation is possible \cite{McLean:2019qcx}. Using the second generation MILC gauge field ensembles at finer lattice spacings, it is possible to simulate bottom quarks close to their physical mass. This allows to simulate all quarks using the same action and by that basically eliminate the systematic uncertainty of the renormalization. In addition they create an array of quark masses filling the range between physical charm and bottom mass and use twisted boundary conditions \cite{Guadagnoli:2005be} to generate data at dedicated values of $q^2$ in the range $q_\text{max}^2\ge q^2\ge 0$. The accessible range of $q^2$ depends on the lattice spacing and the simulated heavy flavor mass.  As can be seen by the colored data points in the upper plots of Fig.~\ref{fig.BsDs}, this increases the $q^2$ directly covered in the numerical lattice simulation as ensembles with finer and finer lattice spacing are used. At a lattice spacing of $a\simeq 0.045$ fm almost the entire range in $q^2$ is covered. As shown in left lower plot, the author of Ref.~\cite{McLean:2019qcx} also claim to have substantially reduced the total uncertainties compared to HPQCD's previous determination \cite{Monahan:2017uby}. To complement the determinations, we also show the updated form factors from the Fermilab/MILC collaboration \cite{Bazavov:2019aom,Bailey:2012rr} in the right plot of Fig.~\ref{fig.BsDs}. In addition Atoui et al.~published results based on ensembles with 2 dynamical flavors \cite{Atoui:2013zza} and preliminary results based on $2+1$ dynamical flavors have been reported in \cite{Flynn:2019jbg,Flynn:2019any}.

The new LHCb measurement \cite{Aaij:2020hsi} determines in combination with \cite{McLean:2019qcx,Monahan:2017uby}
\begin{align}
  |V_{cb}|_\text{CLN} &= [41.4(0.6)(0.9)(1.2)]\times 10^{-3},\\
  |V_{cb}|_\text{BGL} &= [42.3(0.8)(0.9)(1.2)]\times 10^{-3},
\end{align}
where in addition to the BGL $z$ parametrization, the alternative parametrization by Caprini, Lellouch, Neubert (CLN) \cite{Caprini:1997mu} has been used to check for possible systematic effects \cite{Bigi:2016mdz,Bigi:2017njr,Grinstein:2017nlq,Bernlochner:2017xyx,Bernlochner:2019ldg,Gambino:2019sif}. Presently the uncertainties are too large to address the discrepancy between inclusive and exclusive determinations\footnote{For updates see \href{https://hflav.web.cern.ch}{https://hflav.web.cern.ch}} \cite{Amhis:2019ckw}.

\subsection{$B\to D\ell\nu$}\label{Sec.BD}
\begin{figure}[tb]
  \includegraphics[height=0.2\textheight]{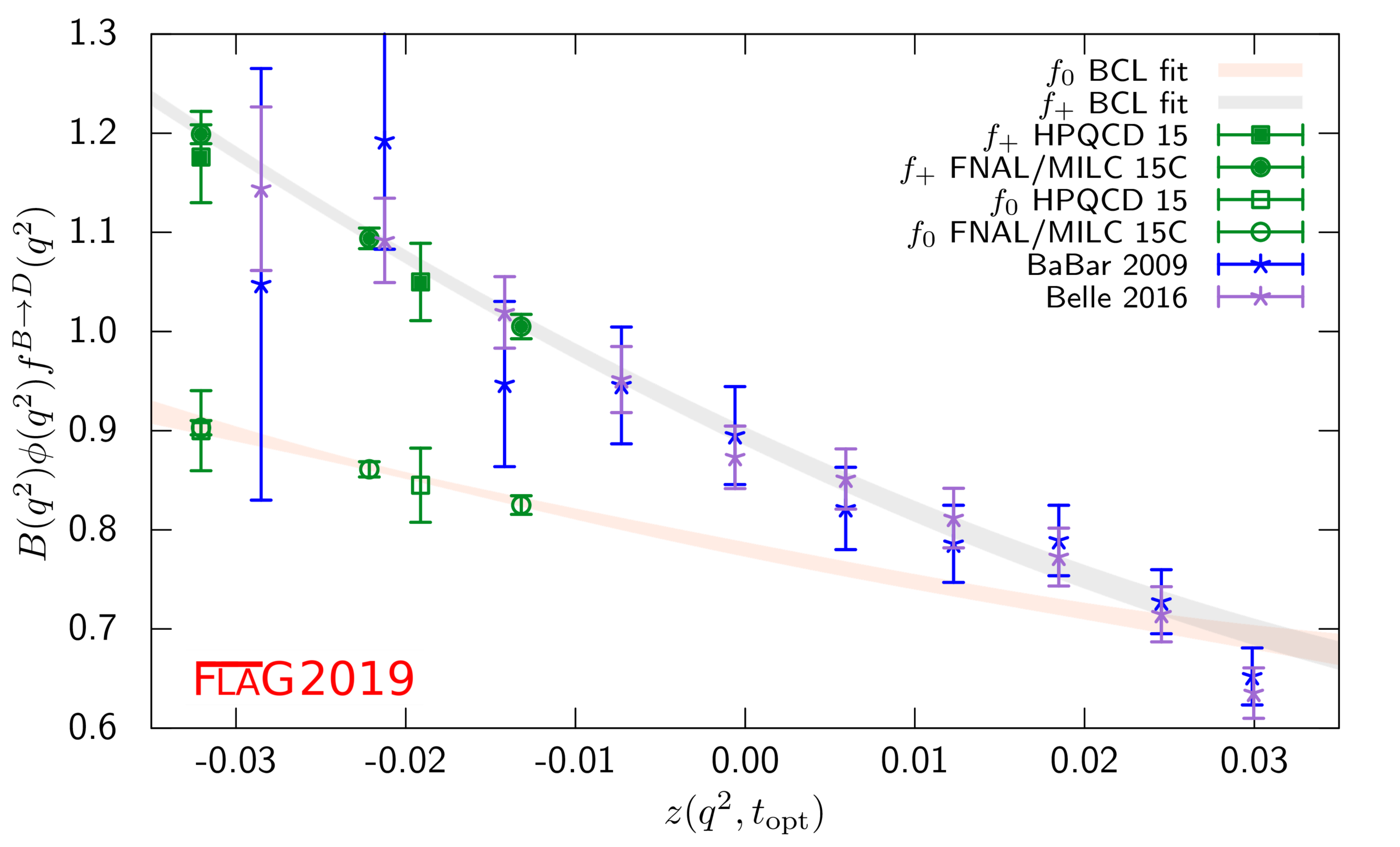}
  \hfill
  \includegraphics[height=0.2\textheight]{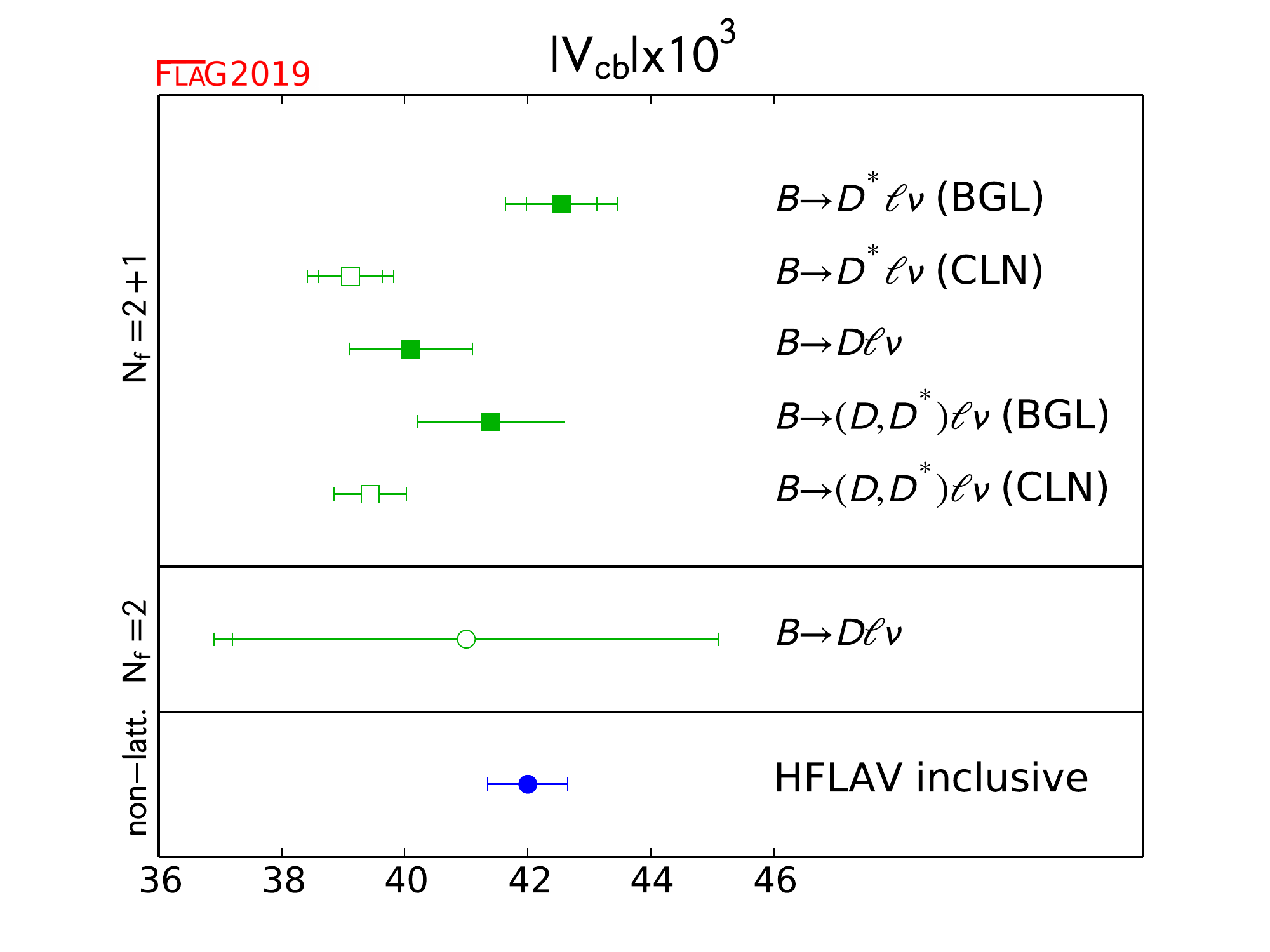}
  \caption{Combined FLAG 2019 analysis \cite{Aoki:2019cca} to determine $|V_{cb}|$ using the lattice determinations of $B\to D\ell\nu$ form factors \cite{Na:2015kha,Lattice:2015rga} and the BaBar \cite{Aubert:2009ac} and Belle \cite{Glattauer:2015teq} experimental data sets.}
  \label{fig.BD}
\end{figure}

Using a charm daughter quark and a spectator with $u/d$ quark mass, we change the diagrams in Fig.~\ref{fig.treelevel} to the determination of $B\to D\ell\nu$ form factors which also allows to extract $|V_{cb}|$. Recently no new results have been published and for completeness we show in Fig.~\ref{fig.BD} the combined FLAG analysis \cite{Aoki:2019cca} based on \cite{Na:2015kha,Lattice:2015rga,Aubert:2009ac,Glattauer:2015teq} extracting
\begin{align}
|V_{cb}|= 40.1(1.0)\times 10^{-3}.
\end{align}

\subsection{$D\to \pi\ell\nu$ and $D\to K\ell \nu$}
Continuing with semi-leptonic decays of $D$-mesons, we study processes in which a charm quark weakly decays to a down or strange quark. Similar to the previously discussed semi-leptonic $B_{(s)}$ decays, these channels allow to extract the CKM matrix elements $|V_{cd}|$ and $|V_{cs}|$. Since leptonic $D_{(s)}$ decays are experimentally well studied, more precise channels to determine  $|V_{cd}|$ and $|V_{cs}|$ exist. Calculating numerically more precise and cheaper 2-point functions on the lattice, the leptonic decay constants $f_D$ and $f_{D_s}$ are typically used to extract the CKM matrix elements. 

\begin{figure}[tb]
  \centering
  \includegraphics[height=0.25\textheight]{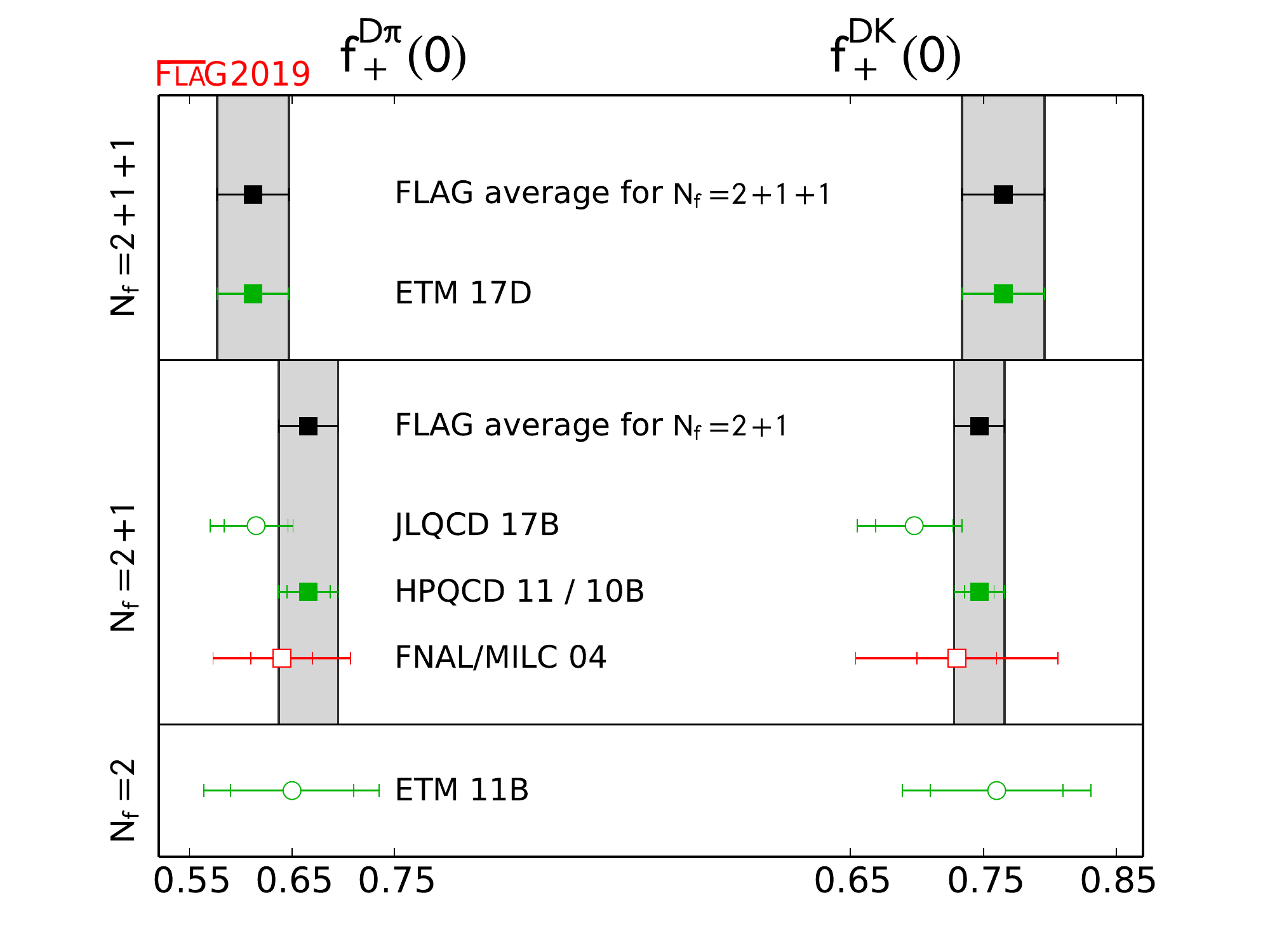}
  \caption{Published results on semi-leptonic $D\to\pi\ell\nu$ and $D\to K\ell\nu$ decays compiled by FLAG \cite{Aoki:2019cca} based on the work by Fermilab/MILC, HPQCD, ETM, and JLQCD \cite{Aubin:2004ej,Na:2010uf,Na:2011mc,Lubicz:2017syv,Kaneko:2017xgg,DiVita:2011py}. }
  \label{fig.Dpi_DK}
\end{figure}

That likely explains why only few lattice calculations for $D$ meson semi-leptonic form factors have been published \cite{Aubin:2004ej,Na:2010uf,Na:2011mc,Lubicz:2017syv}. These decays are kinematically more favorable for a lattice determination than semi-leptonic $B$ decays and the existing calculations focus on directly determining the form factor at zero $q^2$. A summary of existing $D\to\pi\ell\nu$ and $D\to K\ell\nu$ calculations compiled by FLAG \cite{Aoki:2019cca} is shown in Fig.~\ref{fig.Dpi_DK}. Updates on form factor calculations including coverage of the full $q^2$ range have however been reported at recent Lattice conferences \cite{Chakraborty:2019lat,Li:2019phv,Kaneko:2017xgg}.

\section{$R(D^*)$}\label{Sec.vtFF}
Due to observed differences between theoretical predictions and experimental measurements, $R$-ratios testing the universality of lepton flavors in semi-leptonic $B$ decays are in the spotlight of the flavor physics community. Especially the tree-level $B\to D^{(*)}\ell \nu$ decay are in the focus because a global analysis combining results for pseudoscalar and vector hadronic final state reveals a $\sim 3\sigma$ tension \cite{HFLAV:RdRds-2019}. In the case of tree-level decays, $R$-ratios are studied where the branching fraction with $\tau \nu_\tau$ leptonic final state is divided by the branching fraction with $\mu\nu_\mu$ or $e\nu_e$ leptonic final state
\begin{align}
  R^{\tau/l}(D^{(*)})\equiv \frac{BF(B\to D^{(*)}\tau\nu_\tau)}{BF(B\to D^{(*)} l \nu_l)}\qquad\text{with}\quad l=e,\,\mu.
\end{align}
While the difference between electron and muon mass is negligible, depending on the experiment muons or electrons are easier to be identified. The large mass of the tau-leptons however forces a different weight of the two form factors in Eq.~(\ref{Eq.BF}). To extract a theoretical nonperturbative prediction of $R$-ratios from lattice QCD calculations, corresponding form factors over the full $q^2$ range are required. In the case of $B\to D\ell\nu$ decays, we briefly summarized these determinations in Section \ref{Sec.BD}. Combining Fermilab/MILC \cite{Lattice:2015rga} and HPQCD's \cite{Na:2015kha} results, FLAG \cite{Aoki:2019cca} obtains the average
\begin{align}
  R^{\tau/l}(D) = 0.300(8).
\end{align}

\begin{figure}[tb]
  \centering
  \includegraphics[width=0.49\textwidth]{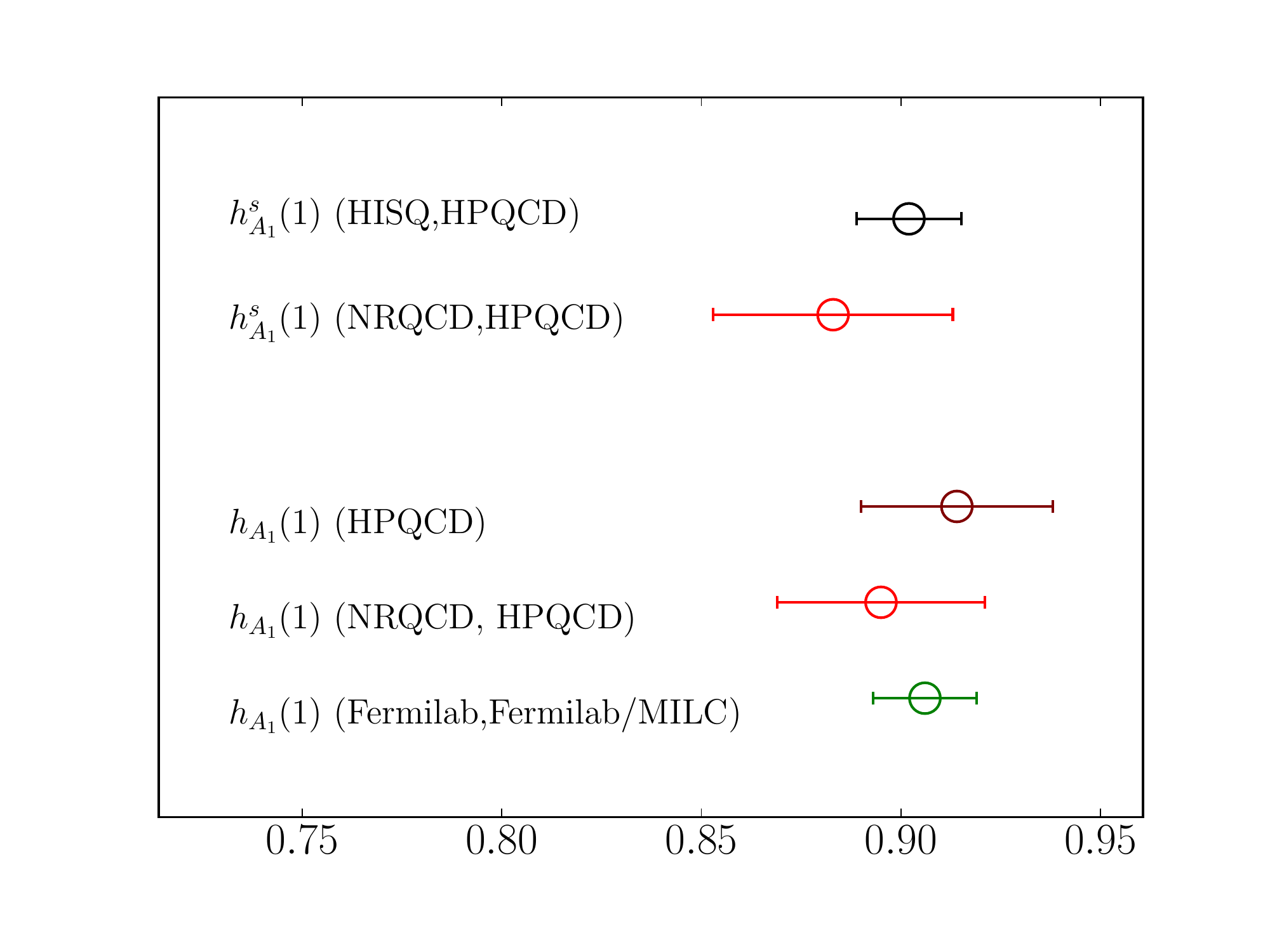}
  \caption{Comparison of different determinations of $h_{A_1}(1)$ for $B\to D^*\ell\nu$ and $h^s_{A_1}(1)$ for $B_s\to D_s^*\ell\nu$ shown by HPQCD in Ref.~\cite{McLean:2019sds} using determinations based on gauge field ensembles with $2+1$ \cite{Bailey:2014tva} or $2+1+1$ dynamical flavors \cite{McLean:2019sds,Harrison:2017fmw}.}
  \label{fig.BDstar_1}
\end{figure}

To date no lattice calculation with full $q^2$ dependence has been published for decays with hadronic vector final state. These decays are conventionally expressed in terms of four form factors 
\begin{align}
  \langle D^*(k,\lambda) |\bar c \gamma_\mu b| B(p)\rangle =& f_V \frac{2i\epsilon_{\mu\nu\rho\sigma}\varepsilon^{*\nu}k^\rho p^\sigma}{M_B+M_D^*},\\        
        \langle D^*(k,\lambda) |\bar c \gamma_\mu\gamma_5 b| B(p)\rangle=& f_{A_0}(q^2)\frac{2M_{D^*} \varepsilon^*\cdot q}{q^2}   q_\mu 
     + f_{A_1} (q^2)(M_{B} + M_{D^*})\left[ \varepsilon^*_{\mu}  - \frac{\varepsilon^*\cdot q}{q^2}   q_\mu \right] \notag\\
    &\quad -f_{A_2}(q^2) \frac{\varepsilon^*\cdot q}{M_{B}+M_{D^*}} \left[ k_\mu + p_\mu - \frac{M_{B}^2 -M_{D^*}^2}{q^2}q_\mu\right],
\end{align}
and considerations are simplified by assuming the narrow width approximation i.e.~treating the vector particle as QCD-stable state and neglecting that it can strongly decay. 
Accounting for the specific kinematics of a heavy bottom to a a heavy charm transition, frequently these form factors are also defined in terms of $w=v_B\cdot v_{D^*}$, the product of the four-velocities $v_B = p_B/M_B$ and $v_{D^*} = k_{D^*}/M_{D^*}$
\begin{align}
  \langle D^*(v_{D^*},\lambda) |\bar c \gamma_\mu b| B(v_B)\rangle =& \sqrt{M_B M_{D^*}} h_V(w) \epsilon_{\mu\nu\rho\sigma}\varepsilon^{*\nu}v^\rho_{D^*} v^\sigma_B,\\        
  \langle D^*(v_{D^*},\lambda) |\bar c \gamma_\mu\gamma_5 b| B(v_B)\rangle=& i \sqrt{M_B M_{D^*}} \notag \\
 &\times \left[ h_{A_1}(w)(1+w) \varepsilon^*_\mu  
     + h_{A_2}(w) \varepsilon^* \cdot v_Bv_{B_\mu} - h_{A_3}(w) \varepsilon^* \cdot v_Bv_{D^*_\mu}\right].
\end{align}  
In this convention the branching fraction is given by
\begin{align}
  \frac{d\Gamma(B\to D^*\ell\nu)}{dw} \approx \frac{G_F^2 M_{D^*}^3}{4\pi^3} \left(M_B-M_{D^*}\right)^2 (w^2-1)^{1/2}|\eta_{EW}|^2|V_{cb}|^2\chi(w) |{\cal F}(w)|^2,
  \label{Eq.BF_BDstarHQ}
\end{align}
where terms proportional to the squared lepton mass have been dropped which is justified in the case of electron or muon leptonic final states. In the limit of zero recoil ($w\to 1$), Eq.~(\ref{Eq.BF_BDstarHQ}) simplifies, $\chi(w) \to 1$ and ${\cal F}(1) = h_{A_1}(1)$. This means only one form factor needs to be determined on the lattice to extract in combination with experimental measurements the CKM matrix element $|V_{cb}|$. At zero recoil lattice determinations of the form factor $h_{A_1}(1)$ exist and the comparison in Fig.~\ref{fig.BDstar_1} also shows that within present uncertainties no dependence on the spectator quark can be resolved.

As highlighted by the introductory remarks on $R^{\tau/l}(D^*)$, obtaining lattice QCD form factors for $B\to D^*\ell\nu$ covering the full range in $q^2$ is of utmost importance. Several collaborations are presently tackling this calculations and have also already reported preliminary results \cite{Aviles-Casco:2019zop,Kaneko:2019vkx,Bhattacharya:2018gan}. Besides the determination of $R^{\tau/l}(D^*)$, lattice QCD form factors with information on the full $q^2$ range will also significantly contribute to the determination of $|V_{cb}^\text{excl}|$. Experimentally, a vector $D^*$ final state is preferred and more measurements exists compared to the pseudoscalar $D$ final state. The determination of $|V_{cb}|$ is however troubled by a long standing $2-3\sigma$ deviation between results based on exclusive and inclusive channels. Lately, the preferred use of the CLN $z$-parametrization for heavy-to-heavy form factor calculations moved into the focus of the discussion \cite{Bigi:2016mdz,Bigi:2017njr}. Whether or not CLN or BGL $z$-parametrizations have an impact on the value of $|V_{cb}^\text{excl}|$, getting additional knowledge on the form factor from a lattice calculation directly covering a large range in $q^2$ will be extremely valuable. Although a more favorable kinematics underlies the recent results for $B_s\to D_s\ell\nu$ decays shown in the upper plots of Fig.~\ref{fig.BsDs}, this setup however clearly demonstrates the potential power of future lattice calculations.

\section{$b$ \& $c$ quark masses}\label{Sec.c_b_mass}

\begin{figure}[tb]
  \includegraphics[height=0.25\textheight]{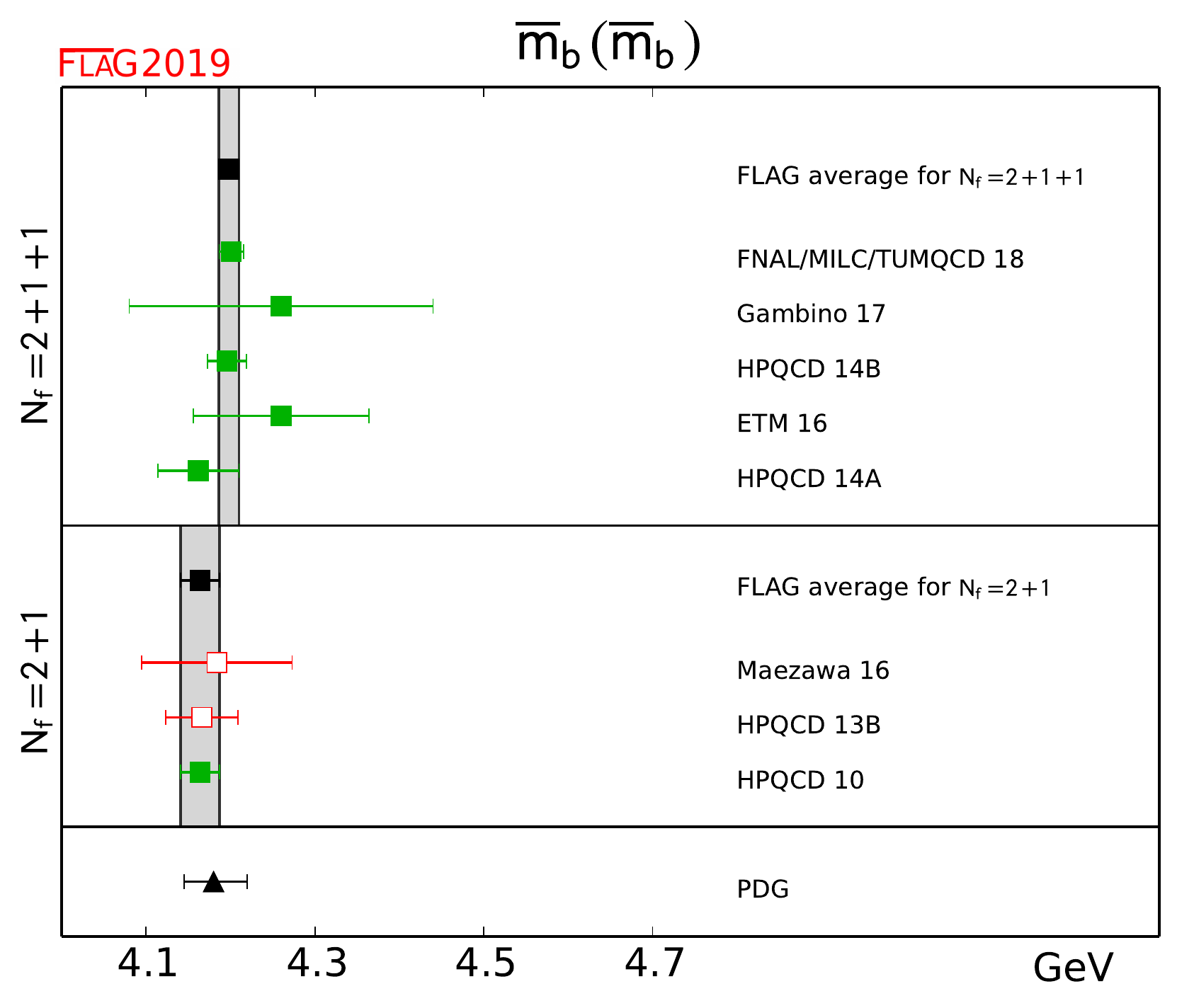}
  \hfill
  \includegraphics[height=0.25\textheight]{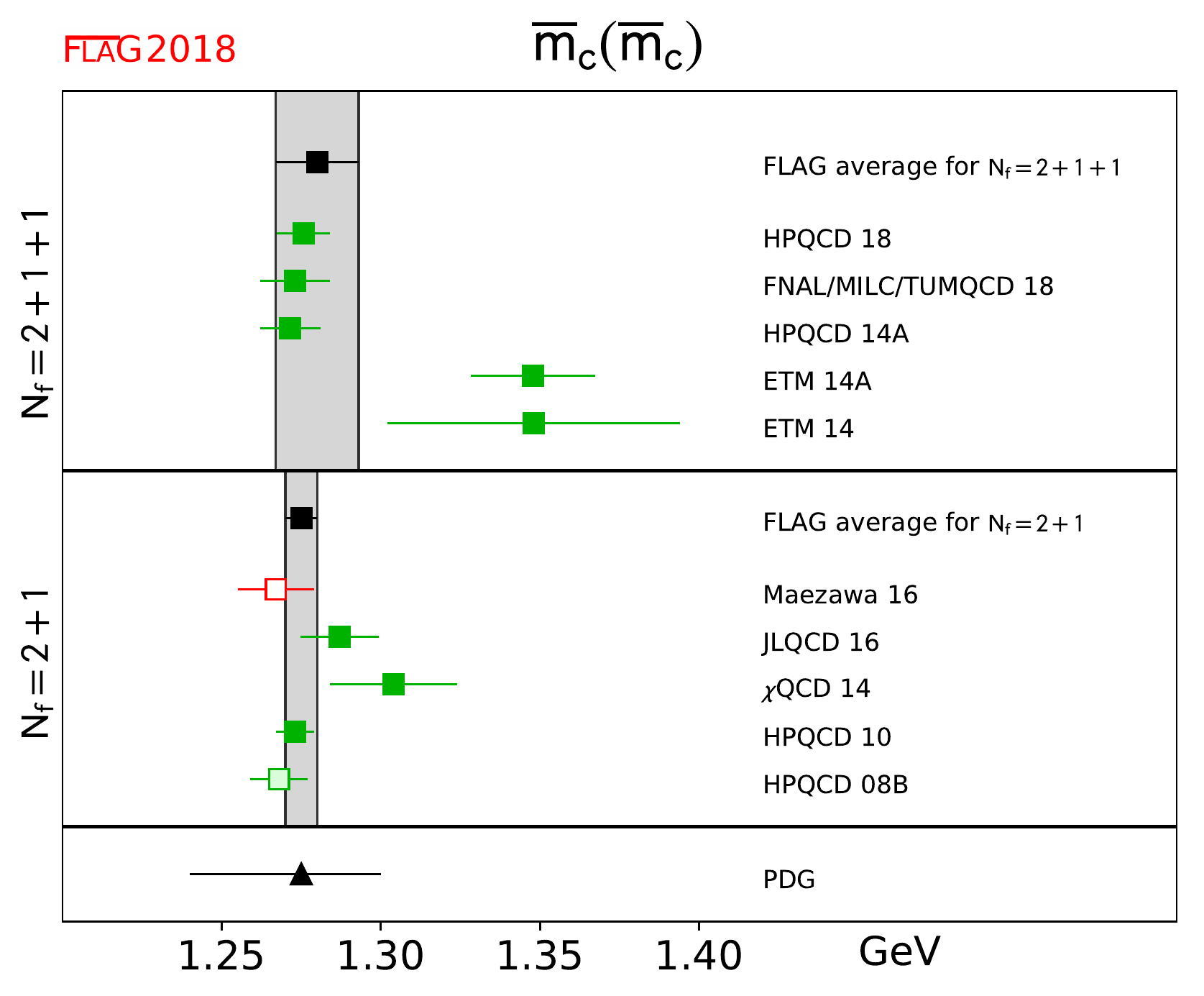}
  \caption{FLAG overview \cite{Aoki:2019cca} of lattice determinations  of the bottom \cite{Bazavov:2018omf,Gambino:2017vkx,Bussone:2016iua,Colquhoun:2014ica,Chakraborty:2014aca,McNeile:2010ji} (left) and charm \cite{Bazavov:2018omf,Chakraborty:2014aca,McNeile:2010ji,Lytle:2018evc,Alexandrou:2014sha,Carrasco:2014cwa,Nakayama:2016atf,Yang:2014sea} (right) quark mass.}
  \label{fig.b_c}
\end{figure}

Figure \ref{fig.b_c} displays an overview of the lattice calculations feeding into the FLAG 2019 \cite{Aoki:2019cca} averages for the determinations of bottom and charm quark masses. All values are presented using the $\overline{\text{MS}}$ scheme and running to the energy scale $\mu$ equal to the respective quark mass. Different methods can be used for the determination. A relatively straightforward method uses the experimental values for $M_\Upsilon$ or $M_{\eta_b}$ to determine the mass of $b$ quarks and correspondingly $M_{D_{(s)}}$ or $M_{\eta_c}$ for the mass of $c$ quarks. An alternative method proceeds by determining on the lattice the Euclidean time moments of pseudoscalar-pseudoscalar correlators for heavy-quark currents which is followed by an operator product expansion in perturbative QCD.

Presently the most precise determination of the bottom quark mass is obtained by Fermilab/MILC/TUMQCD \cite{Bazavov:2018omf} which is based on the 2+1+1 flavor HISQ ensembles and uses a sophisticated fit strategy based on HQET, HMrAS$\chi$PT, and Symanzik effective theory. All determinations of the bottom quark mass are mutually consistent. The FLAG averages separated by determinations based on gauge field configurations with 2+1 and 2+1+1 dynamical flavor are
\begin{align}
  2+1:&\quad \overline{m}_b(\overline{m}_b)=4.164(23) \gev,\\
  2+1+1:&\quad \overline{m}_b(\overline{m}_b)=4.198(12) \mev.
\end{align}

The situation is somewhat different in the case of charm quark mass determinations. Both HPQCD \cite{Chakraborty:2014aca,Lytle:2018evc} and Fermilab/MILC/TUMQCD \cite{Bazavov:2018omf} determined the charm quark mass very precisely. Their determinations agree with each other and dominate the FLAG average. There is however a tension to the determinations by the ETM collaboration \cite{Alexandrou:2014sha,Carrasco:2014cwa} which is also based on ensembles with 2+1+1 dynamical flavors but finds larger values for the  charm quark mass. The FLAG averages again separated by used gauge field ensembles with 2+1 and 2+1+1 flavors are
\begin{align}
 2+1:&\quad \overline{m}_c(\overline{m}_c)=1.275(5) \gev,\\
 2+1+1:&\quad \overline{m}_c(\overline{m}_c)=1.280(13) \gev.
\end{align}

\section{Neutral $B$ meson mixing}\label{Sec.mixing}

\begin{figure}[tb]
  \centering
  \begin{picture}(69,45)(0,2)
    \put(4,30){\includegraphics[width=0.38\textwidth]{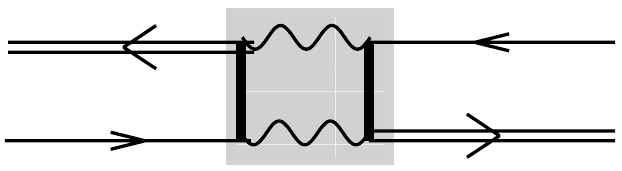}}
    \put(1,36){\large $B$}
    \put(62,36){\large $\overline{B}$}
    \put(15,43){{$\bar b$}}
    \put(53,43){{$\bar d$}}    
    \put(15,28){{$d$}}
    \put(53,29){{$b$}}
    \put(32,44){$W$}
    \put(32,29){$W$}    
    \put(28,36){$t$}
    \put(36,36){$t$}    
    
    \put(4,5){\includegraphics[width=0.38\textwidth]{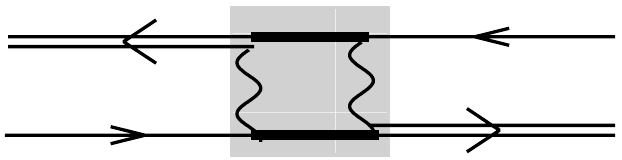}}    
    \put(1,11){\large $B$}
    \put(62,11){\large $\overline{B}$}
    \put(15,18){{$\bar b$}}
    \put(53,18){{$\bar d$}}    
    \put(15,3){{$d$}}
    \put(53,4){{$b$}}    
    \put(33,18){$t$}
    \put(33,4){$t$}    
    \put(28,11){$W$}
    \put(33.5,11){$W$}
  \end{picture}
  \hspace{12mm}
  \begin{picture}(60,45)(0,2)
    \put(0,12){\includegraphics[width=65mm]{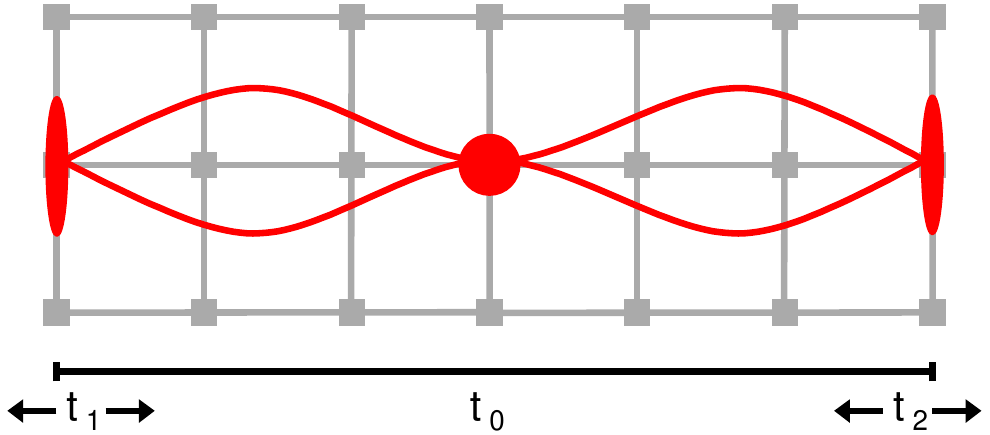}}
  \end{picture}
  \caption{Left: Sketch of the box diagrams showing the dominant short distance contribution to neutral $B$ meson mixing. Right: Schematic quark line diagram of the lattice computation to calculate the nonperturbative contributions to neutral meson mixing.}
  \label{fig.mixing}
\end{figure}
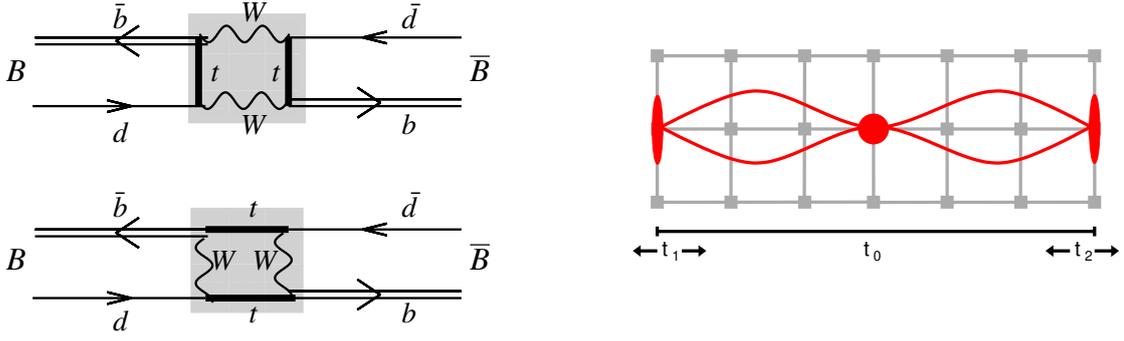

Although a loop-level process in the SM, neutral $B_{(s)}$ meson mixing provides important tests on the SM and, furthermore, opens a window to study top physics and determine the CKM matrix elements $|V_{td}|$ and $|V_{ts}|$. Neutral meson mixing occurs in the SM by the box diagrams sketched on the left in Fig.~\ref{fig.mixing}. Since in the case of $B_{(s)}$ mesons, the top quark contribution in the loop dominates, the process is short distance and well suited for a lattice calculation. Experimentally, neutral meson mixing is observed by measuring extremely precisely the oscillation frequency $\Delta m_q$ for $q=d,\,s$. Conventionally $\Delta m_q$ is parametrized by
\begin{align}
          \Delta m_q = \frac{G_F^2m^2_W}{6\pi^2} \eta_B S_0 M_{B_q} {f_{B_q}^2B_{B_q}} \abs{V_{tq}^*V_{tb}}^2,
\end{align}
where in addition to perturbatively calculated functions, the nonperturbative contributions given by the decay constant $f_{B_q}$ and the bag parameter $B_{B_q}$ enter.  In particular favorable for a lattice determination is to consider the ratio of  $B_s$ meson mixing over $B$ meson mixing \cite{Bernard:1998dg}
\begin{align}
  \frac{\Delta m_s}{\Delta m_d} = \frac{M_{B_s}}{M_{B_d}}\,{\xi^2} \, \frac{\abs{V_{ts}}^2}{\abs{V_{td}}^2}\qquad \text{with}\qquad
  {\xi^2} = \frac{f^2_{B_s} B_{B_s}}{f^2_{B_d}B_{B_d}}.
\end{align}
Perturbative calculated contribution cancels and the measured ratio of oscillation frequencies is proportional to the ratio of the meson masses times the ratio of the CKM matrix elements $|V_{ts}|^2/|V_{td}|^2$ times $\xi^2$, the nonperturbative contributions. The lattice determination of $\xi$ has the advantage that in a ratio also many lattice uncertainties cancel. To obtain $\xi$, we need to calculate decay constants $f_{B_q}$ and bag parameters $B_{B_q}$ on the lattice which are related to the matrix elements
\begin{align}
    \langle 0|\bar q \gamma^\mu\gamma_5 b |B_q(p)\rangle &= i f_{B_q} p^\mu_{D_q}, \label{Eq.fB}\\
    \langle \bar B_q^0 | [\bar b \gamma^\mu(1-\gamma_5) q][\bar b \gamma_\mu(1-\gamma_5) q] |B_q^0\rangle &= \frac{3}{8} f^2_{B_q} M_{B_q}^2 B_{B_q}. \label{Eq.BB}
\end{align}
Equation (\ref{Eq.fB}) is a simple vacuum-to-meson 2-point function and lattice determinations of $f_{B_{(s)}}$ can be found in \cite{Bussone:2016iua,Bazavov:2017lyh,Hughes:2017spc,Dowdall:2013tga,Christ:2014uea,Aoki:2014nga,Na:2012kp,McNeile:2011ng,Bazavov:2011aa,Bernardoni:2014fva,Carrasco:2013zta}, whereas Eq.~(\ref{Eq.BB}) requires to evaluate a 3-point function schematically shown on the right of Fig.~\ref{fig.mixing}. The matrix element describes how a $B$ meson changes to a $\overline{B}$ meson. The mixing occurs in a point-like operator connecting four quark lines. One possible implementation is to keep the four-quark operator fixed at time $t_0$ and vary the times $t_1$ and $t_2$ where in each case two of the four quark lines starting at $t_0$ are contracted to create the $B$ and $\overline{B}$ meson, respectively.

\begin{figure}[tb]
  \centering
  \includegraphics[height=0.193\textheight]{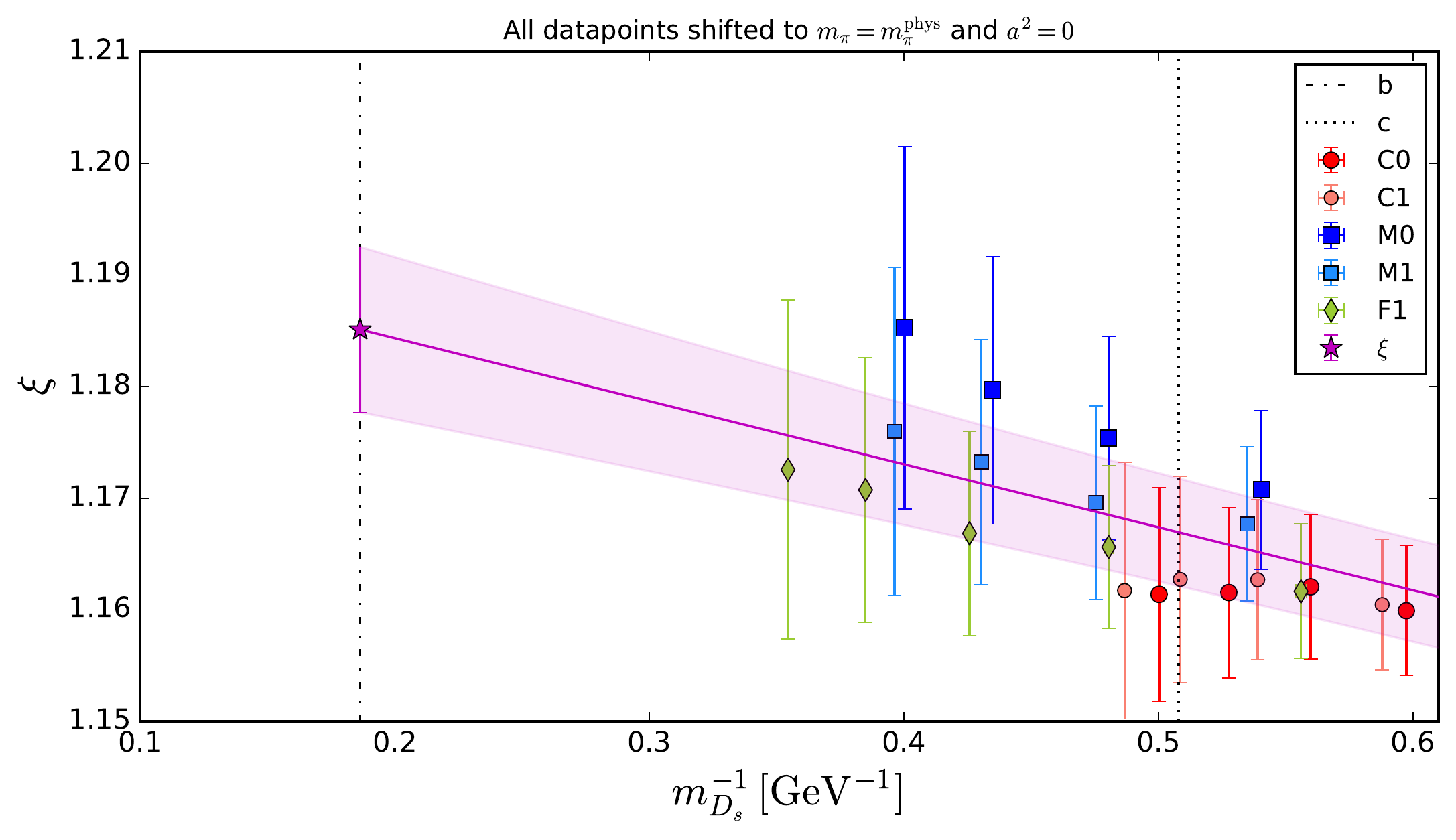}
  \includegraphics[height=0.193\textheight]{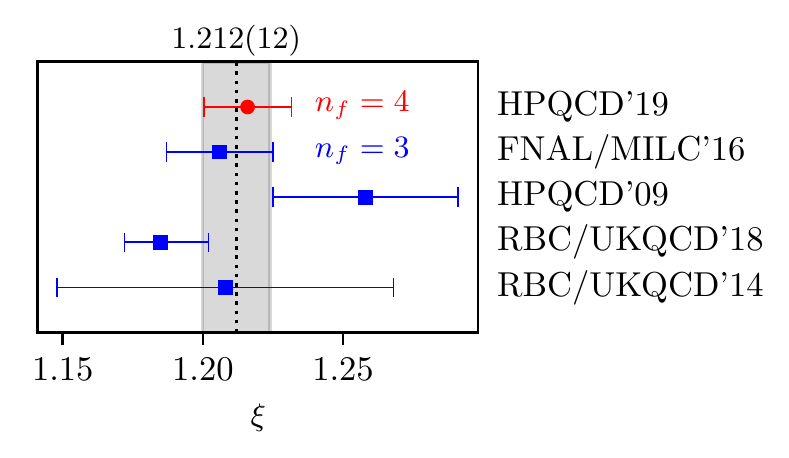}
  \caption{Left: Demonstration of RBC/UKQCD's new approach to use heavy M\"obius domain wall fermions to determine $\xi$ \cite{Boyle:2018knm}. By simulating an array of charm-like heavy flavor masses, a linear extrapolation to the $b$-quark mass (vertical dash-dotted line) is enabled. Right: Average of $\xi$ determinations compiled by HPQCD \cite{Dowdall:2019bea} and based on Refs.~\cite{Boyle:2018knm,Dowdall:2019bea,Bazavov:2016nty,Aoki:2014nga,Albertus:2010nm,Gamiz:2009ku}}
  \label{fig.xi}
\end{figure}

As mentioned above, many uncertainties cancel in the ratio $\xi$. Hence $\xi$ is a good quantity to explore new concepts for the calculation of heavy-light matrix elements. On the left of Fig.~\ref{fig.xi} we demonstrate RBC/UKQCD's use of M\"obius domain wall fermions for the determination of $\xi$ \cite{Boyle:2018knm}. The calculation is performed using different values of the heavy flavor mass to measure on the lattice an array of heavy-light and heavy-strange mixing matrix elements. This results in a sequence of values for $\xi$ which slightly grow when increasing the heavy flavor mass. The overlapping symbols with different colors correspond to measurements performed at different values of the lattice spacing and indicate that discretization effects are small. The almost linear dependence of $\xi$ on the heavy flavor mass, allows to extrapolate the simulated data to the $b$ quark mass and read-off the physical value of $\xi$. A similar concept, the heavy HISQ discretization, is used by HPQCD to obtain the first determination of $\xi$ based on 2+1+1 flavor gauge field ensembles \cite{Dowdall:2019bea}. A comparison of the different determinations exhibiting overall good agreement is shown on the right of Fig.~\ref{fig.xi}.

\begin{figure}[tb]
  \centering
  \includegraphics[height=0.20\textheight]{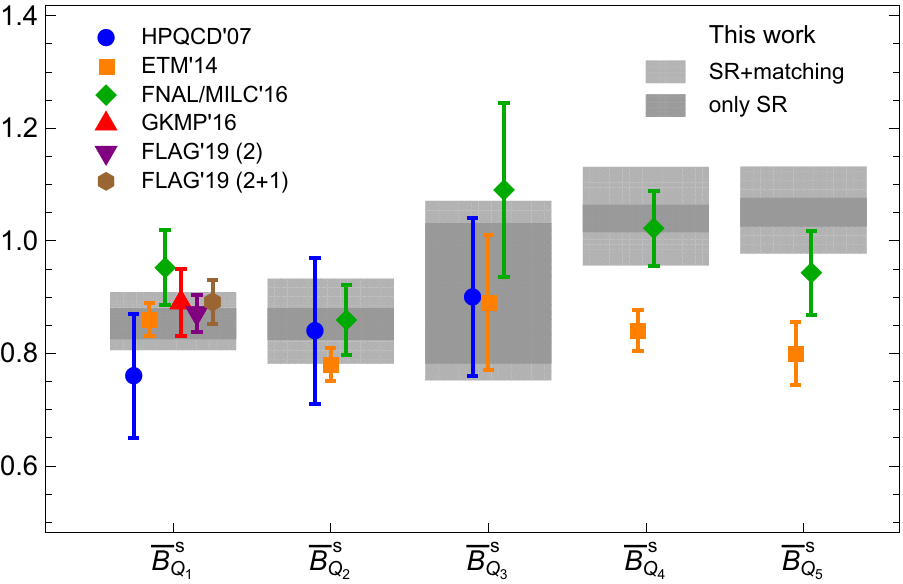}
  \hspace{15mm}  
  \includegraphics[height=0.25\textheight]{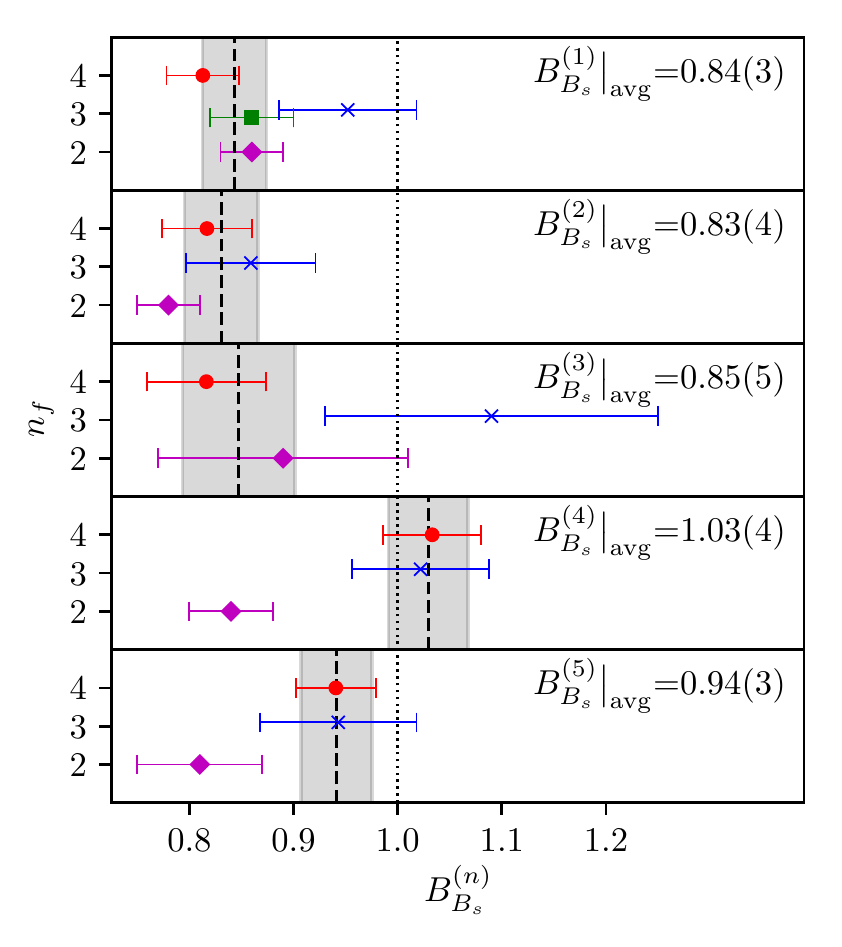}
  \caption{Comparison and averages for all five $B_s$ meson mixing operators. The left shows the comparison performed by King et al.~\cite{King:2019lal} showing predictions obtained from QCD sum rules \cite{King:2019lal,Grozin:2016uqy} as well as lattice calculations \cite{Carrasco:2013zta,Bazavov:2016nty, Gamiz:2009ku} and FLAG averages \cite{Aoki:2019cca}. The plot on the right presented by HPQCD \cite{Dowdall:2019bea} compares their new result (red circles) based on 2+1+1 flavor gauge field ensembles  \cite{Dowdall:2019bea} to other lattice determinations.  Blue crosses and green square determinations refer to results based on 2+1 flavor gauge field ensembles by Fermilab/MILC's \cite{Bazavov:2016nty} and HPQCD \cite{Gamiz:2009ku}, the purple diamonds refers to ETM collaborations determination based on 2 flavor gauge field ensembles \cite{Carrasco:2013zta}. Black dashed lines with gray bands indicate averages of results based on 2+1 and 2+1+1 flavor gauge field configurations.}
  \label{fig.Bbag}
\end{figure}

\begin{figure}[tb]
  \centering
  \includegraphics[height=0.25\textheight]{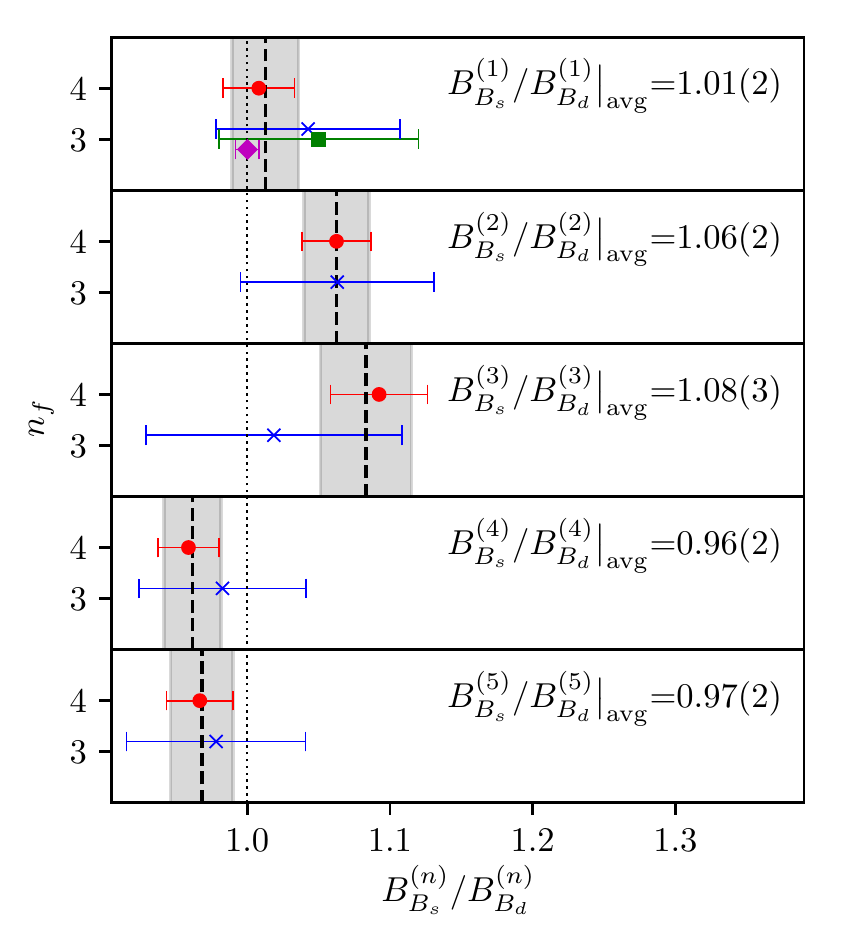}    
  \hspace{10mm}
  \includegraphics[height=0.23\textheight]{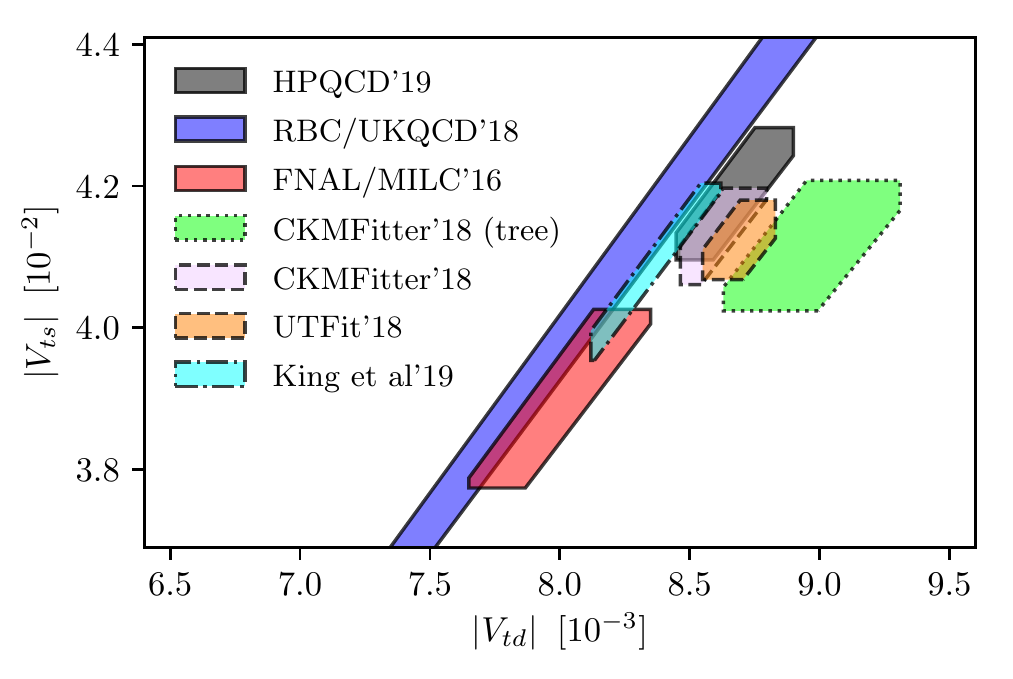}  
  \caption{Left: Comparison and averages for the ratio of $B_s$ over $B$ meson bag parameters for all five mixing operators presented by HPQCD \cite{Dowdall:2019bea}.  Red circles denote HPQCD's determination based on 2+1+1 flavor gauge field ensembles \cite{Dowdall:2019bea}, blue crosses and green square determinations are based on 2+1 flavor gauge field ensembles by Fermilab/MILC's \cite{Bazavov:2016nty} and HPQCD \cite{Gamiz:2009ku}, the purple diamonds refers to RBC/UKQCD's value obtained on 2+1 flavor gauge field configurations \cite{Boyle:2018knm}. Black dashed lines with gray bands indicate averages. Right: Constraints of the CKM matrix elements $|V_{td}|$ and $|V_{ts}|$ derived from  $B_{(s)}$-meson mixing (solid boundary) \cite{Boyle:2018knm,Dowdall:2019bea,Bazavov:2016nty,King:2019lal} , unitarity constraints of SM (dashed boundary) \cite{Charles:2004jd, CKMfitter,Bona:2006ah, UTfit} or unitarity triangle fit (dotted boundary) \cite{Charles:2004jd, CKMfitter}.}
  \label{fig.VtdVts}
\end{figure}

However, not only the ratio $\xi$ but also the individual bag parameters are important phenomenological quantities to constrain the SM. Since neutral meson mixing occurs at 1-loop in the SM, it is particularly sensitive to new physics contributions. While in the SM model, only the operator in Eq.~(\ref{Eq.BB}) contributes, four other local operators may be significant for extensions of the SM. This so called SUSY basis defines in addition
\begin{align}
  &Q_2^q =\left[\bar b(1-\gamma_5)q \right]\left[\bar b(1-\gamma_5)q \right], \qquad
  &Q_3^q =\left[\bar b^\alpha(1-\gamma_5)q^\beta \right]\left[\bar b^\beta (1-\gamma_5)q^\alpha \right], \notag \\ 
  &Q_4^q =\left[\bar b(1-\gamma_5)q \right]\left[\bar b(1+\gamma_5)q \right], \qquad
  &Q_5^q =\left[\bar b^\alpha (1-\gamma_5)q^\beta \right]\left[\bar b^\beta(1+\gamma_5)q^\alpha \right],  
\end{align}
where $\alpha,\,\beta$ are color indices shown explicitly for contractions across the two bilinears. Besides contributing to meson mixing in BSM theories, some of these operators also contribute to the width difference $\Delta \Gamma_q$ in the SM.\footnote{For pioneering work to determine $\Delta \Gamma_s$ including dimension-7 operators see Ref.~\cite{Davies:2019gnp}.} Figure \ref{fig.Bbag} shows the outcome of determining $B_s$ meson bag parameters for all five operators. The comparison on the left includes results based on QCD sum rules \cite{King:2019lal,Grozin:2016uqy} and different lattice calculations \cite{Carrasco:2013zta,Bazavov:2016nty, Gamiz:2009ku}. The comparison on the right adds the first calculation on 2+1+1 flavor gauge field configurations by HQCD \cite{Dowdall:2019bea}. The discrepancy in determinations for operators $Q_4^s$ and $Q_5^s$ might be related to different intermediate renormalization schemes \cite{Boyle:2017ssm}. Moreover, a comparison of bag parameter ratios $B_{B_s}/B_{B_d}$ is shown in the left plot of Fig.~\ref{fig.VtdVts}.

Using the information obtained from the ratio $\xi$ or the bag parameters, constraints on the CKM matrix elements $|V_{td}|$ and $|V_{ts}|$ can be obtained and put in relation to constraints based on unitarity of the SM. The present status is shown in the right plot in Fig.~\ref{fig.VtdVts}.

\section{Summary}
Lattice QCD provides a well-established framework to calculate nonperturbative contributions of the strong force to phenomenologically relevant quantities. Although calculations often turn into multi-year projects, the recent years have seen quite significant progress improving technical aspects of simulations determining heavy-light quantities. With these new techniques at hand, the near future looks promising for many important updates on quantities presented here. The references pointing to preliminary results mostly presented at recent Lattice conference provide details of upcoming calculations. Furthermore, it is important to highlight that new developments also target the calculation of quantities so far not tackled on the lattice. Conceptual new ideas have e.g.~been brought forward to calculate inclusive semi-leptonic decays on the lattice \cite{Hashimoto:2017wqo,Hansen:2017mnd,Bailas:2020cho} or to study radiative decays \cite{Kane:2019jtj,Sachrajda:2019uhh}.

In addition there are many more process than covered in this overview. Some of which have already been calculated on the lattice in the past, others are presently in progress. To name only a few examples: semi-leptonic decays with flavor changing neutral currents \cite{Horgan:2013pva,Horgan:2013hoa,Horgan:2015vla,Bouchard:2013pna,Bailey:2015dka,Flynn:2016vej}, $B_c$ decays and $R(J/\psi)$ \cite{Cooper:2019byi,Colquhoun:2016osw}, or exclusive baryonic decays \cite{Detmold:2015aaa,Datta:2017aue,Meinel:2016dqj}.

\section*{Acknowledgments}
The author thanks his colleagues at the University of Colorado Boulder as well as his RBC and UKQCD collaborators for helpful discussions and suggestions. OW acknowledges support from DOE grant DE--SC0010005.

{\small
\bibliography{../B_meson}
\bibliographystyle{JHEP-notitle}
}
\end{document}